\pdfoutput=1
\documentclass[times,twocolumn,final]{elsarticle}

 \RequirePackage{geometry}
 \global\bibsep=0pt

\usepackage{framed,multirow}

\usepackage{amssymb}
\usepackage{latexsym}
\usepackage{mathtools}

\usepackage{url}
\usepackage{xcolor}

\usepackage{hyperref}

\usepackage{makecell}
\usepackage{longtable}
\usepackage{balance}
\usepackage{subfig}

\newcommand{\tdtd}{3-D$\rightarrow$2-D }
\newcommand{\rtdtd}{2-D$\rightarrow$3-D }
\newcommand{\argmax}{\mathop{\mathrm{argmax}}}

\journal{arXiv}

\begin{document}

\begin{frontmatter}

\title{DISCOVER: 2-D Multiview Summarization of Optical Coherence Tomography Angiography for Automatic Diabetic Retinopathy Diagnosis}%

\author[1,2]{Mostafa El Habib Daho}
\author[1,2]{Yihao Li}
\author[1,2]{Rachid Zeghlache}
\author[3,4]{Hugo Le Boité}
\author[5]{Pierre Deman}
\author[6]{Laurent Borderie}
\author[7]{Hugang Ren}
\author[7]{Niranchana Mannivanan}
\author[4]{Capucine Lepicard}
\author[1,8,2]{Béatrice Cochener}
\author[4]{Aude Couturier}
\author[9,4]{Ramin Tadayoni}
\author[10,2]{Pierre-Henri Conze}
\author[1,2]{Mathieu Lamard}
\author[2]{Gwenolé Quellec\corref{cor1}}
\cortext[cor1]{LaTIM - IBRBS - 22, avenue Camille Desmoulins - 29200 Brest, France - Tel.: +33-6-12-25-22-57}
\ead{gwenole.quellec@inserm.fr}

\address[1]{Univ Bretagne Occidentale, Brest, F-29200 France}
\address[2]{Inserm, UMR 1101, Brest, F-29200 France}
\address[3]{Sorbonne University, Paris, F-75006 France}
\address[4]{Service d'Ophtalmologie, H\^{o}pital Lariboisi\`ere, APHP, Paris, F-75475 France}
\address[5]{ADCIS, Saint-Contest, F-14280 France}
\address[6]{Evolucare Technologies, Le Pecq, F-78230 France}
\address[7]{Carl Zeiss Meditec, Dublin, CA 94568, USA}
\address[8]{Service d'Ophtalmologie, CHRU Brest, Brest, F-29200 France}
\address[9]{Paris Cit\'e University, Paris, F-75006 France}
\address[10]{IMT Atlantique, Brest, F-29200 France}


\begin{abstract}
Diabetic Retinopathy (DR), an ocular complication of diabetes, is a leading cause of blindness worldwide.
Traditionally, DR is monitored using Color Fundus Photography (CFP), a widespread 2-D imaging modality. However, DR classifications based on CFP have poor predictive power, resulting in suboptimal DR management. Optical Coherence Tomography Angiography (OCTA) is a recent 3-D imaging modality offering enhanced structural and functional information (blood flow) with a wider field of view. This paper investigates automatic DR severity assessment using 3-D OCTA. A straightforward solution to this task is a 3-D neural network classifier. However, 3-D architectures have numerous parameters and typically require many training samples. A lighter solution consists in using 2-D neural network classifiers processing 2-D en-face (or frontal) projections and/or 2-D cross-sectional slices. Such an approach mimics the way ophthalmologists analyze OCTA acquisitions: 1) en-face flow maps are often used to detect avascular zones and neovascularization, and 2) cross-sectional slices are commonly analyzed to detect macular edemas, for instance. However, arbitrary data reduction or selection might result in information loss. Two complementary strategies are thus proposed to optimally summarize OCTA volumes with 2-D images: 1) a parametric en-face projection optimized through deep learning and 2) a cross-sectional slice selection process controlled through gradient-based attribution. The full summarization and DR classification pipeline is trained from end to end. The automatic 2-D summary can be displayed in a viewer or printed in a report to support the decision. We show that the proposed 2-D summarization and classification pipeline outperforms direct 3-D classification with the advantage of improved interpretability.
\end{abstract}


\end{frontmatter}


\section{Introduction}

Diabetic Retinopathy (DR), a complication of diabetes, is a major and growing cause of vision impairment and blindness. By 2040, around 600 million people throughout the world will have diabetes \citep{ogurtsova_idf_2017}, a third of whom will have DR \citep{yau_global_2012}. One major problem in the management of DR is its reliance on an older imaging technique, namely Color Fundus Photography (CFP). Various classifications based on CFP were proposed over the years \citep{david_airlie_1969, etdrs_research_group_fundus_1991, wilkinson_proposed_2003}. Unfortunately, decisions based on these classifications have poor predictive power. For instance, a severe non-proliferative DR case evolves to proliferative complication in 51.5\% of cases, with only 17.1\% evolving to a high risk of blindness \citep{etdrs_research_group_fundus_1991}. This makes the management of DR challenging: clinicians often err on the side of caution and treat all those patients to mitigate the risk of complications. In the past decades, a significant number of studies have relied on CFP images for the automatic assessment of DR. The application of machine learning techniques, particularly deep learning, to these images has shown promising results in the detection and categorization of DR \citep{ting_dl_ophta_2019, quellec_explain_2021}. While these advancements are noteworthy, such techniques will always suffer from the poor predictive power of CFP. Fortunately, new imaging modalities are emerging that may improve predictions.

Optical Coherence Tomography (OCT) is a non-invasive imaging technique that uses interferometric information of partially coherent light to create cross-sectional (2-D B-scans) and three-dimensional (C-scans) structural images of biological tissues. Within optical scattering media, it can penetrate a few millimeters in depth, with micrometer resolution, and is therefore particularly well-suited to image the retina \citep{hitzenberger_three-dimensional_2003}. OCT Angiography (OCTA) is a motion-sensitive extension of OCT enabled by fast OCT acquisitions: it was shown to contrast the retinal vasculature and can provide quantitative blood flow information \citep{baumann_total_2011}. An OCTA acquisition can be summarized by two volumes: a \textit{structure volume}, obtained by averaging consecutive 3-D scans, and a \textit{flow volume}, describing the amplitude of local intensity variations across those consecutive 3-D scans \citep{gorczynska_comparison_2016}. To analyze the blood flow in specific vascular plexuses, clinicians generally inspect en-face (or frontal) Maximal Intensity Projections (MIP) of the flow volume in the corresponding retinal or choroidal layers \citep{gorczynska_comparison_2016}. In such 2-D projections, each 1-D A-scan is replaced with the maximal intensity value (throughout the entire A-scan or within the considered layers only). Recent OCTA devices enable ultra-widefield acquisitions of the retina (90$^{\circ}$) \citep{niederleithner_ultra-widefield_2023}. In summary, OCTA can capture ultra-widefield volumetric structural and functional (flow) images of the retina: it is, therefore, a promising technique to diagnose various ocular pathologies \citep{alam_supervised_2019, yang_deep_2023, xu_av-casnet_2023, zang_deep-learningaided_2023}. In particular, DR can clearly benefit from OCTA \citep{yang_classification_2022}: 1) the structure volume allows objective and quantitative assessment of diabetic macular edema (DME), 2) flow MIPs allow quantification of retinal vascular plexuses, non-perfusion and vessel density as well as the identification of damage; \citet{vujosevic_standardization_2021} lists the various biomarkers of DR and DME in OCTA acquisitions.

Computer-aided DR diagnosis using OCTA is an emerging field of research: it is motivated by the above promises (i.e., useful biomarkers) and the challenge of integrating large amounts of data (i.e., 3-D ultra-widefield structural and flow images). In particular, various quantitative metrics were automated to assist in early detection, staging, and progression of DR \cite{sun_optical_2021}. Those metrics quantify retinal fluid volumes \citep{guo_automated_2020}, retinal vasculature features (e.g., density, tortuosity) \citep{alam_supervised_2019, lo_federated_2021, khalili_pour_automated_2023}, avascular zones \citep{guo_automatic_2019, guo_automated_2020}, including the Foveal Avascular Zone (FAZ) \citep{alam_supervised_2019, li_diagnosing_2022}, and proliferative DR features such as neovascularization \citep{vaz-pereira_update_2021}.

Through a radiomics approach, these features were used for automatic DR severity assessment \citep{carrera-escale_radiomics-based_2023, ryu_deep_2022, khalili_pour_automated_2023}. Various methods were also investigated to assess DR severity directly from OCTA images. Some authors classified 2-D en-face MIP images with 2-D Convolutional Neural Networks (CNN): \citet{le_transfer_2020, andreeva_dr_2020, lo_federated_2021, ryu_deep_2021, ryu_deep_2022} classified one en-face flow MIP (superficial plexus, deep plexus or full retina), \citet{heisler_ensemble_2020} jointly classified two en-face flow MIPs (superficial and deep plexus) and the corresponding en-face structure MIPs, \citet{yasser_automated_2022} and \citet{li_diagnosing_2022} jointly classified en-face flow MIPs and 2-D feature maps derived from feature segmentation. Other authors classified 3-D images with 3-D CNNs: \citet{zang_diabetic_2022} classified one 2-channel (structure and flow) 3-D image, \citet{li_multimodal_2022} jointly classified one 2-channel 3-D images and one 2-D Line-Scanning Ophthalmoscope (LSO) localizer, and \citet{li_3-d_2023} jointly classified two 2-channel 3-D images acquired with different fields of view (6$\times$6mm$^2$ and 15$\times$15mm$^2$).

Theoretically, the radiomics approach and the 2-D classification approach are suboptimal in the sense that relevant features useful for classification may have been lost during the preprocessing (MIP) and feature extraction steps. However, the 3-D classification approach also has some limitations. First, compared to their 3-D counterparts, 2-D neural architectures have better pre-trained weights (e.g., ImageNet weights) and have fewer parameters to optimize, thus requiring fewer training samples. Given the limited OCTA data sets available, this aspect is critical. Second, end-to-end 3-D classification lacks the interpretability power of the radiomics approach and, to a lesser extent, of the end-to-end 2-D classification approach. To alleviate these limitations, we propose an end-to-end 3-D image classification approach relying on 2-D views as intermediate steps, the 2-D view extraction process being trainable. This guarantees that: 1) 2-D neural architectures can be used at the end of the classification pipeline, while 2) relevant problem-specific features can be extracted at the beginning of the pipeline. In detail, two types of view are extracted:
\begin{enumerate}
  \item 2-D en-face (or frontal) projections, generalizing the flow (or structure) MIP images used to assess OCTA flow features \citep{yang_classification_2022},
  \item selected 2-D slices (B-scans), often used to assess structural OCT features \citep{yang_classification_2022}: attribution methods \citep{sundararajan_axiomatic_2017, samek_evaluating_2017} are used to identify the most relevant scan lines in the en-face projection, which are deemed worthy of further investigation in 2-D.
\end{enumerate}
To maximize the interpretability of en-face projections, a novel ``model dropout'' mechanism is introduced. Projections are processed by an ensemble of 2-D image classifiers, and during training, classifiers in the ensemble are dropped randomly. The extracted 2-D features thus become more classifier-independent, i.e., more general and hopefully more meaningful to the human eye.

This paper aims to automate the most recent DR severity classification \citep{wilkinson_proposed_2003} from OCTA acquisitions, using the proposed interpretable classification approach, one step toward better management of DR.

\section{Related Methods}

\subsection{\tdtd Projection}

In recent years, \tdtd projection was used to solve various medical image analysis tasks. One such task is 2-D/3-D registration: in that case, the 2-D image is registered to a 2-D projection of the 3-D image. \citet{fei_automatic_2006} used that strategy to register 2-D X-ray images to 3-D Computed Tomography (CT) images: the goal was to compare the ability of both modalities to detect cardiac calcifications. More recently, \citet{schaffert_metric-driven_2018} and \citet{jaganathan_self-supervised_2023} registered intra-operative 2-D X-rays to pre-operative 3-D CT images for navigation purposes during minimally invasive surgeries. Alternatively, \citet{van_houtte_deep_2022} registered intra-operative 2-D X-rays to a pre-operative 3-D atlas. For 2-D/3-D registration, \tdtd projection is typically achieved by simulating a 2-D perspective projection from the 3-D image to the plane: in \citet{schaffert_metric-driven_2018} and \citet{jaganathan_self-supervised_2023}, a linear system of equations (point-to-plane correspondences) is solved; \citet{van_houtte_deep_2022} relies on projective spatial transformers. In \citet{jaganathan_self-supervised_2023}, an adversarial domain adaptation step is added.

\tdtd projection was also used for en-face segmentation of 3-D images. This is useful when ground truth masks are obtained using en-face projections of the 3-D images or using a different 2-D imaging modality. \citet{li_image_2020} segmented blood vessels and foveal avascular zones in 3-D structural and flow OCTA images: the ground truth masks were obtained using the flow MIP between the Internal Limiting Membrane (ILM) layer and the Outer Plexiform Layer (OPL). \citet{lachinov_projective_2021} segmented geographic atrophies, a sign of age-related macular degeneration, in 3-D structural OCT images: the ground truth masks were obtained using 2D Fundus Auto Fluorescence (FAF) images. They also segmented blood vessels in 3-D OCT: the ground truth masks were obtained using en-face OCT images. \citet{le_novel_2021} segmented blood vessels in 3-D Adaptive Optics OCT (AO-OCT) images: ground truth masks were obtained using en-face AO-OCT images. For en-face segmentation of 3-D images, variations on the U-Net architecture are typically used: in the proposed architectures, the encoder part contains 3-D operations, and the decoder part contains 2-D operations \citep{li_image_2020, lachinov_projective_2021}. Alternatively, an LSTM is combined with the encoder part of a 2-D U-Net in \citet{le_novel_2021}.

A variation on the previous task is the generation of high-quality 2-D images from 3-D images for visualization purposes or, optionally, for downstream segmentation tasks. \citet{forsgren_high-throughput_2022} generated high-quality 2-D projections from low-quality 3-D fluorescence microscopy images, which can be acquired fast. In biology, \citet{haertter_deepprojection_2022} projected curved 2-D manifolds from 3-D microscopy image stacks on a 2-D plane. Solutions to this problem are supervised: \citet{haertter_deepprojection_2022} use a U-Net-like structure, \citet{forsgren_high-throughput_2022} use conditional GANs with a U-Net-like backbone.

\tdtd projection was also used for fast feature detection or segmentation in 3-D images. In this case, ground truth masks are obtained from 3-D images: 2-D projections are simply used as intermediate steps. \citet{shen_efficient_2021} detected 3-D junction points from various tree-like structures (blood vessels, neurons) in 2-D projections, followed by \rtdtd reverse mapping. Similarly, \citet{wang_vc-net_2021} segmented 3-D microvessels in 2-D projections of 3-D brain Magnetic Resonance Imaging (MRI), followed by \rtdtd reverse mapping. For this task, non-parametric \tdtd projections are used, namely MIP \citep{shen_efficient_2021, wang_vc-net_2021}. Finally, \citep{guo_end--end_2021} segmented blood vessels in OCTA images. A variation on MIP was used in that study: all voxel intensities in the same retinal layer are multiplied by a layer-specific trainable weight prior to MIP.

A final task explored in this paper is 3-D image classification. \citet{statsenko_deep_2022} use \tdtd projection to diagnose COVID-19-associated pneumonia in 3-D CT images. \citet{gupta_performance_2020} concatenates multiple 2-D projections of coronary arteries/branches to identify diseased coronary arteries/branches in 3-D Computed Tomography Angiography (CTA). \citet{mandal_computer-aided_2023} use \tdtd projection to differentiate lentigo maligna from atypical intraepidermal melanocytic proliferation, two melanoma subtypes, in 3-D Reflectance Confocal Microscopy (RCM). The motivation is to take advantage of 2-D neural architectures: compared to their 3-D counterparts, 2-D architectures have better pre-trained weights and fewer parameters to optimize \citep{statsenko_deep_2022, gupta_performance_2020}. Using 2-D architectures also reduces computational requirements \citep{mandal_computer-aided_2023}. For this task, non-parametric \tdtd projections are also used: averaged 2-D slices \citep{statsenko_deep_2022}, MIP \citep{gupta_performance_2020, mandal_computer-aided_2023}.

Unlike previous 3-D image classification tasks, we propose a trainable parametric \tdtd projection to allow the extraction of discriminant and interpretable problem-specific features. The proposed approach is more general than \citep{guo_end--end_2021}'s solution for segmentation, which solely emphasizes specific retinal layers. Other trainable techniques proposed for 1) 2-D/3-D registration, 2) en-face segmentation of 3-D images, or 3) high-quality 2-D image generation assume that dense ground truth (e.g., 2-D images) is available to supervise \tdtd projection training. This is not our case: the only supervision signal at our disposal is the DR diagnosis. This calls for a different architecture to ensure that feature localization is not lost in the en-face plane, as demonstrated in this paper.

\subsection{Attribution Methods}
\label{sec:attribution}

Over the past decade, with the growing popularity of deep learning, various methods were introduced to visualize what ``black box'' CNNs have learned. The purpose was notably to attribute an importance score to each image pixel for a given output target (e.g., a class prediction or a neuron activation). They can be classified into gradient-based or perturbation-based attribution methods.

The simplest gradient-based attribution method is the Saliency method by \citet{simonyan_deep_2014}, which computes the gradient of the output target with respect to each image pixel. This approach is advantageous because it allows pixel-level attribution at the cost of a single gradient backpropagation. However, not all operations in a CNN are invertible, which often leads to unsatisfactory attributions. \citet{zeiler_visualizing_2014} thus introduced a mechanism to correctly backpropagate attributions through MaxPool operations (Deconvolution method), and \citet{springenberg_striving_2015} introduced another way to correctly backpropagate them through ReLU activations (Guided Backprop method). These ideas were extended in Layer-Wise Relevance Propagation (LRP) and Deep Learning Important FeaTures (DeepLIFT), with additional properties: LRP ensures that the magnitude of any output is conserved through the backpropagation process \citep{bach_pixel-wise_2015}; DeepLIFT bases the attributions on the difference between the activation of each neuron (for the current image) to its ``reference activation'' \citep{shrikumar_learning_2017}.

The simplest perturbation-based attribution method is the occlusion method by \citet{zeiler_visualizing_2014}, which studies how the output target is impacted when parts of the image (square patches) are occluded. \citet{ribeiro_why_2016} introduced Local Interpretable Model-agnostic Explanations (LIME), which looks for superpixels with the strongest association with the output target by successively turning superpixels off and on. Finally, Integrated Gradients can be regarded as both a gradient-based and a perturbation-based attribution method: 1) a set of images is generated by multiplying all pixel intensities in the target image by a constant factor ranging from 0 (the first image or reference) to 1 (the last image or target); 2) for each image in the series, a gradient-based attribution score is computed; 3) the final attribution is the integral of gradients over the series \citep{sundararajan_axiomatic_2017}. Intuitively, the most relevant features are detected early in the series and have, therefore, a higher integral. Unlike the other gradient-based attribution methods, multiple gradient backpropagation operations are needed. More generally, all perturbation-based attribution methods require multiple CNN evaluations, implying higher computation times and memory requirements.

To the best of our knowledge, this study is the first attempt to use attribution methods to extract and analyze relevant 2-D slices in a 3-D volume.

\begin{figure}[!t]
    \centering
    \includegraphics[width=.35\textwidth]{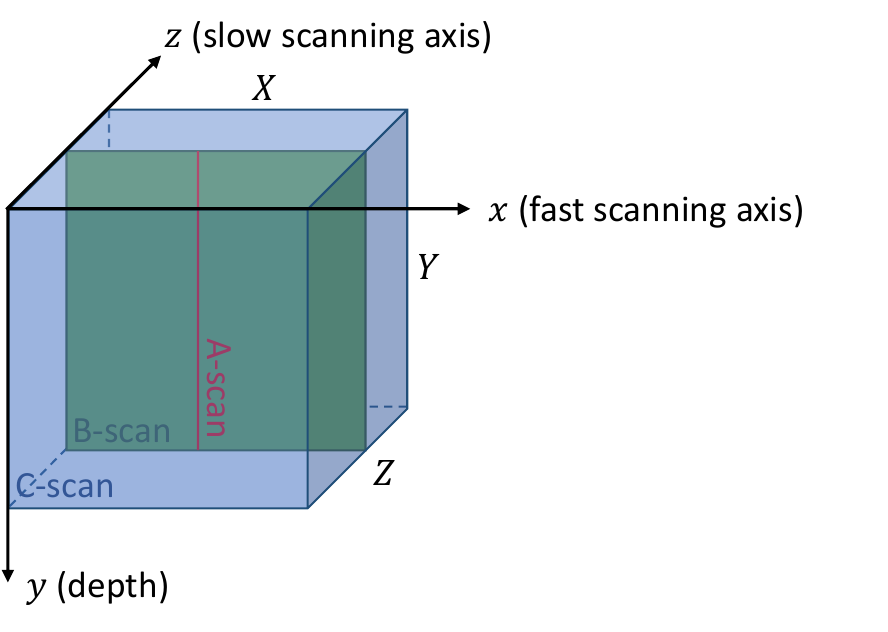}
    \caption{Geometry of an Optical Coherence Tomography (OCT) acquisition. A 3-D B-scan consists of multiple 2-D B-scans, which in turn consist of multiple 1-D A-scans.}
    \label{fig:geometry}
\end{figure}

\begin{figure*}[!t]
    \centering
    \includegraphics[width=\textwidth]{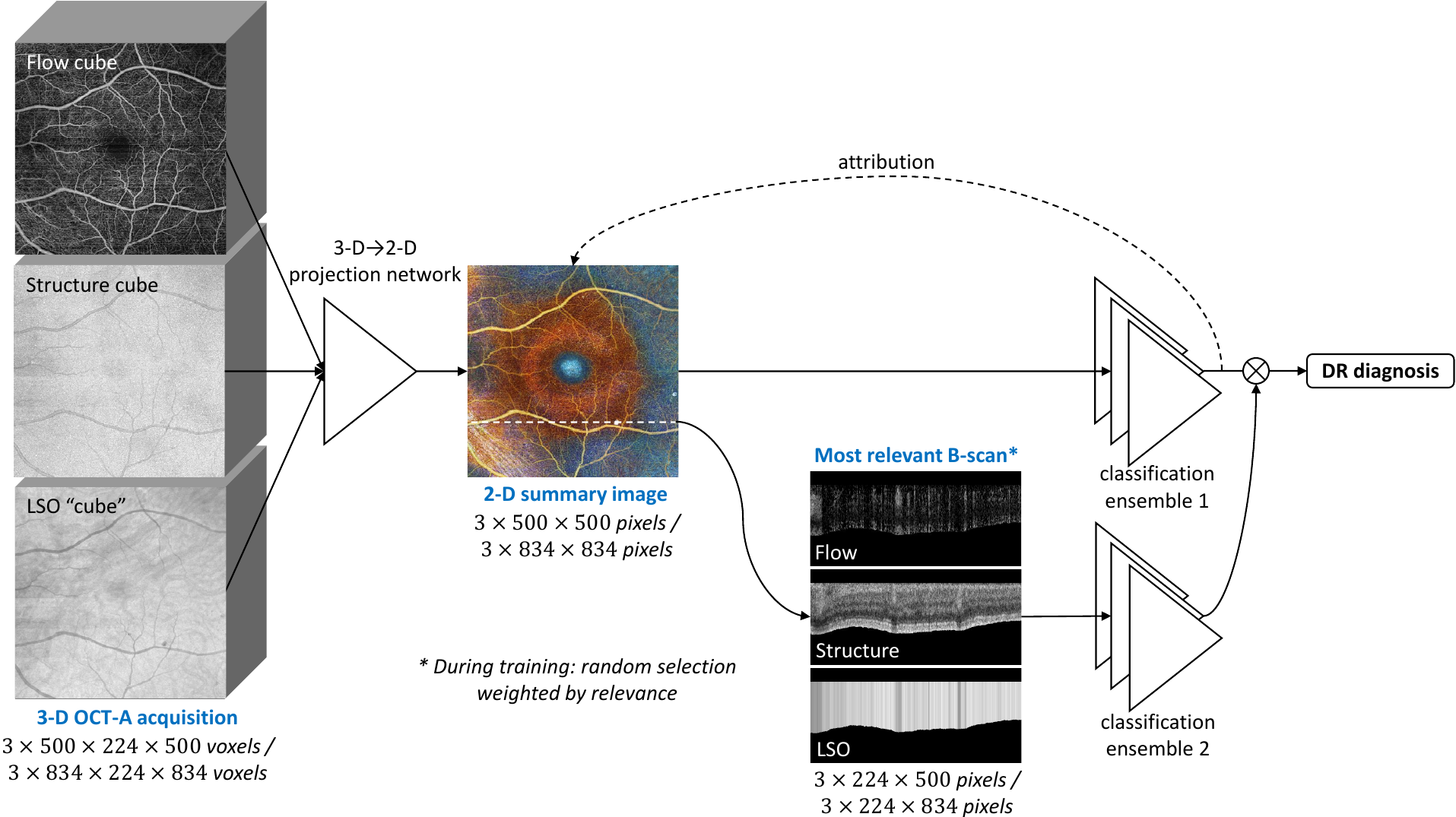}
    \caption{Overview of the proposed approach. A multi-channel 3-D volume is summarized as a 2-D image through a \tdtd projection network, detailed in Fig. \ref{fig:projector}. Next, a first classification branch classifies this 2-D summary image in order to produce a DR diagnosis. Through an attribution method, the most relevant 2-D B-scans are selected. Then, a second classification branch classifies the selected B-scans to improve the DR diagnosis. Each classification branch is an ensemble of classifiers, detailed in Fig. \ref{fig:ensemble}. In this figure, each 3-D channel in the input volume is represented by its Maximum Intensity Projection (MIP).}
    \label{fig:overview}
\end{figure*}

\section{Proposed Method}

\subsection{Overview and Notations}
\label{sec:overview}

The geometry of an OCT/OCTA acquisition is illustrated in Fig.~\ref{fig:geometry}. Let $y$ denote the depth axis along which the partial coherent light penetrates the tissues (A-scan). Let $x$ denote the fast scanning axis: a B-scan is thus indexed by $x$ and $y$. Let $z$ denote the slow scanning axis: a volume (C-scan) is thus indexed by $x$, $y$, and $z$ and en-face projections by $x$ and $z$. Let $X \times Y \times Z$ denote the size of the C-scans in voxels. Finally, a multi-channel volume (e.g., with a flow and a structure channel) is indexed by $c$, $x$, $y$, and $z$.

Given $I$, a preprocessed multi-channel OCTA acquisition (see section \ref{sec:preprocessing}), and $N$ acquisition-level labels, the goal is to predict whether or not experts would assign the $n$-th label to acquisition $I$, $n = 1..N$. In the experiments (see section \ref{sec:experiments}), the goal is to assess DR severity in the patient's eye, according to the 5-level ICDR scale \citep{wilkinson_proposed_2003}: no DR (level 0), mild non-proliferative DR (NPDR, level 1), moderate NPDR (level 2), severe NPDR (level 3), proliferative DR (PDR, level 4); DR severity assessment is formulated as an $N$-label classification problem ($N=4$): is DR severity $d(I)$ greater than or equal to level $n$? Let $\mathbf{p}(I) = \left\lbrace p_n(I) = p(d(I) \geq n) \in [0; 1], n=1..N \right\rbrace$ denote the probabilistic predictions and let $\lambda_n(I) \in \{0, 1\}$, $n=1..N$, denote the ground truth labels.

As illustrated in Fig. \ref{fig:overview}, the preprocessed 3-D acquisition $I$ is converted to a 2-D summary image $\Pi(I)$, defined as a parametric en-face projection of $I$ (see details in section \ref{sec:projection}). $\Pi(I)$ is defined as a color (3-channel) image for two reasons:
\begin{enumerate}
    \item Interpretability: it can be displayed in a viewer or inserted in a report for human inspection.
    \item Compatibility with off-the-shelf 2-D neural architectures with ImageNet pre-trained weights.
\end{enumerate}

Next, $\Pi(I)$ is classified by a first ensemble $C_1$ of 2-D off-the-shelf image classifiers, as described in section \ref{sec:classification_projection}. A first estimation $\mathbf{p}^{(1)}(I)$ of the probabilistic prediction $\mathbf{p}(I)$ is given by $C_1 \circ \Pi(I)$.

Then, based on attributions derived from $\mathbf{p}^{(1)}(I)$, the most relevant B-scans $S(I)$ are selected, as described in section \ref{sec:bscan_selection}. For interpretability purposes, the number of selected B-scans is limited to $N$: one per classification output. The motivations are:
\begin{enumerate}
    \item A small number of B-scans is compatible with human inspection in a viewer or a report.
    \item Each selected B-scan is associated with a severity cutoff and can, therefore, be used to document the course of action associated with that cutoff (treatment, follow-up, etc.).
\end{enumerate}

Finally, the selected B-scans $S(I)$ are classified by a second ensemble $C_2$ of 2-D off-the-shelf image classifiers: a second estimation $\mathbf{p}^{(2)}(I)$ of the probabilistic prediction is given by $C_2 \circ S(I)$. It is combined with $\mathbf{p}^{(1)}(I)$ to obtain the final probabilistic prediction $\mathbf{p}(I)$, as described in section \ref{sec:second_final_classification}.

\subsection{Preprocessing}
\label{sec:preprocessing}

\begin{figure}[!t]
    \centering
    \includegraphics[width=0.43\textwidth]{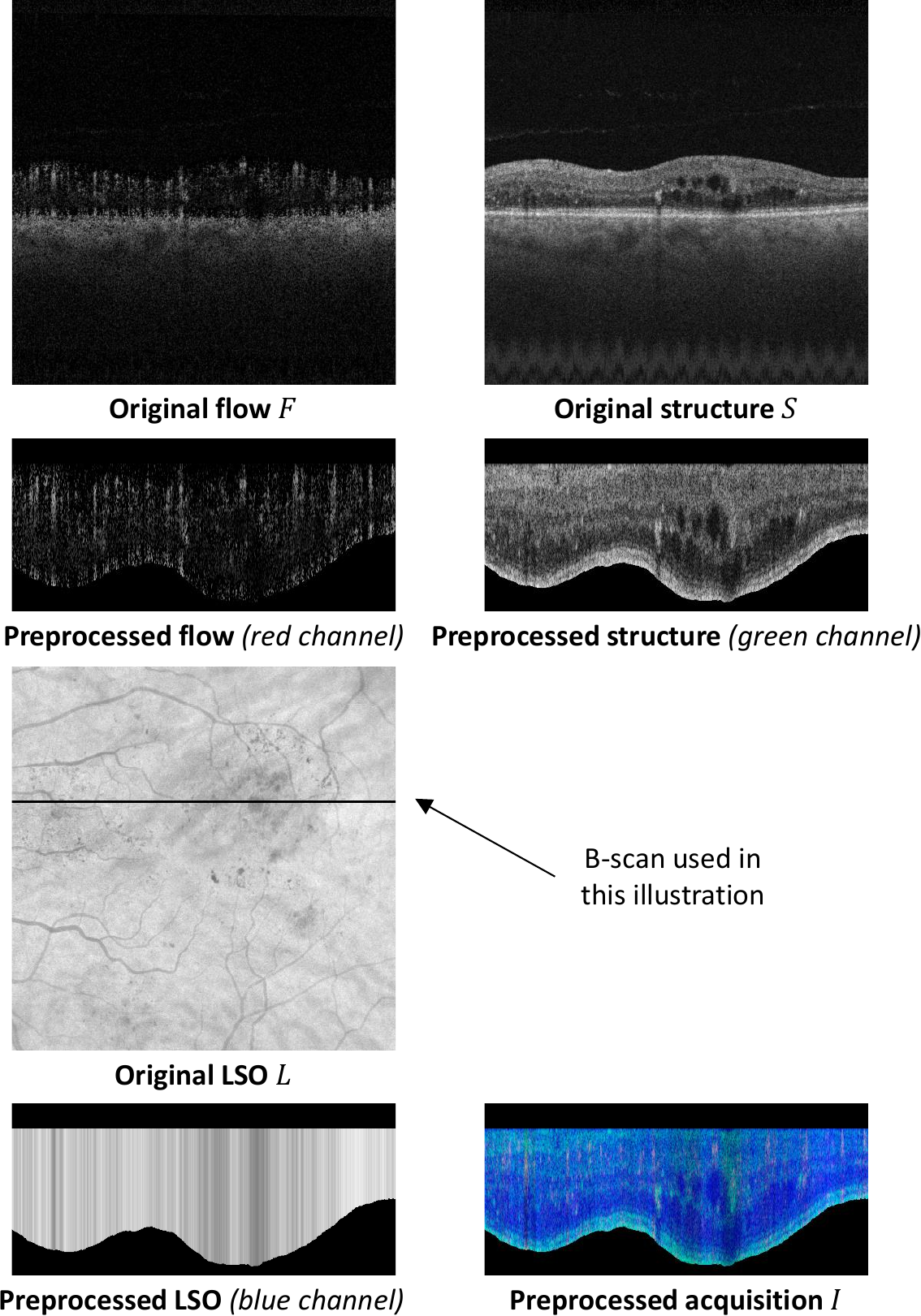}
    \caption{Preprocessing pipeline for OCTA acquisitions (see section \ref{sec:preprocessing}) illustrated on one B-scan. Each original 2-D flow and structure B-scan is flattened, masked out, and cropped. The original 2-D LSO en-face localizer is transformed into a 3-D volume by duplicating pixel intensities along the depth axis (within the masked region). A 3-channel 3-D volume is obtained by stacking the resulting three volumes (flow, structure, LSO).}
    \label{fig:preprocessing}
\end{figure}

An OCTA acquisition is stored as two volumes:
\begin{enumerate}
    \item a structure volume $S$, where the retinal layers and various retinal anomalies (e.g., fluid) are visible, among other structures (e.g., the choroid, below the retina, and the vitreous core, above it),
    \item a flow volume $F$, where the blood vessels of the retina and the choroid are particularly highlighted.
\end{enumerate}
Additionally, OCTA acquisitions are usually associated with:
\begin{enumerate}
    \item A 2-D en-face localizer $l$, aligned with the OCTA data (size: $X \times Z$ pixels), to track eye motion. The PLEX Elite 9000 (Carl Zeiss Meditec Inc. Dublin, California, USA) device, for instance, is associated with a Line Scanning Ophthalmoscope (LSO) subsystem for that purpose \citep{niederleithner_ultra-widefield_2023}.
    \item Automatically-segmented surfaces delineating the vitreoretinal interface, namely the Inner Limiting Membrane (ILM), and the chorioretinal interface, below the Retinal Pigment Epithelium (RPE). Let $s_{sup}$ and $s_{inf}$ denote those two surfaces, respectively. They are stored as matrices of $X \times Z$ pixels: $s(x, z)$ represents the depth of surface $s$ in the $(x, z)$ A-scan of volumes $S$ or $F$.
\end{enumerate}
Although the LSO image is used primarily for motion tracking, it offers a complementary view on the retina (different optical properties, different wavelength, etc.), analogous to a grayscale fundus image \citep{niederleithner_ultra-widefield_2023}. We propose to analyze it jointly with the OCTA data. Therefore, we propose to preprocess an OCTA acquisition as follows (see Fig. \ref{fig:preprocessing}).

First, an ``LSO volume'' $L$ is created by duplicating the LSO localizer along the y-axis: $L(x, y, z) = l(x, z), \forall x, y, z$.

Second, a mask volume $M$ of $X \times Y \times Z$ voxels is created: $M(x, y, z) = 1$ if $s_{inf}(x, z) \leq y \leq s_{sup}(x, z)$, 0 otherwise, $\forall x, y, z$. The flow, structure, and LSO volumes are multiplied by $M$, element-wise, to mask the choroid and vitreous core out. Let $I'$ denote the 3-channel volume:
\begin{equation}
    I' = [F \odot M, S \odot M, L \odot M] \;\; .
\end{equation}

Third, the retinal region is flattened by shifting all voxels of $I'$ along the y-axis, so that the ILM surface is set to a small constant depth $y = Y_0$. Let $I''$ denote the resulting volume:
\begin{equation}
    I''(c, x, y + Y_0, z) = I'(c, x, y + s_{inf}(x, z), z), \forall c, x, y \in [0; Y - Y_0), z \; .
\end{equation}
Parameter $Y_0$ is set to a non-zero value $(0 < Y_0 < Y)$ to limit the loss of useful information during random data augmentation (see section \ref{sec:second_classification}). Its value is determined as described in section \ref{sec:hyperopt}. This flattening process ensures that all the relevant information is concentrated at the top of $I''$. The advantage of flattening the ILM surface, instead of the RPE ($s_{inf}$), is that its segmentation is less error-prone, due to a better contrast. This reduces the risk of alignment errors, which would lead to discontinuities between neighboring A-scans.

Fourth, $I''$ is cropped: all voxels with a depth $y > Y_1$ are discarded, $Y_0 \leq Y_1 \leq Y$. Parameter $Y_1$ is chosen to ensure the retinal region is never occluded. The final preprocessed image $I$ denotes the cropped version of $I''$.

\subsection{\tdtd Projection}
\label{sec:projection}

\begin{figure*}[!t]
    \centering
    \includegraphics[width=\textwidth]{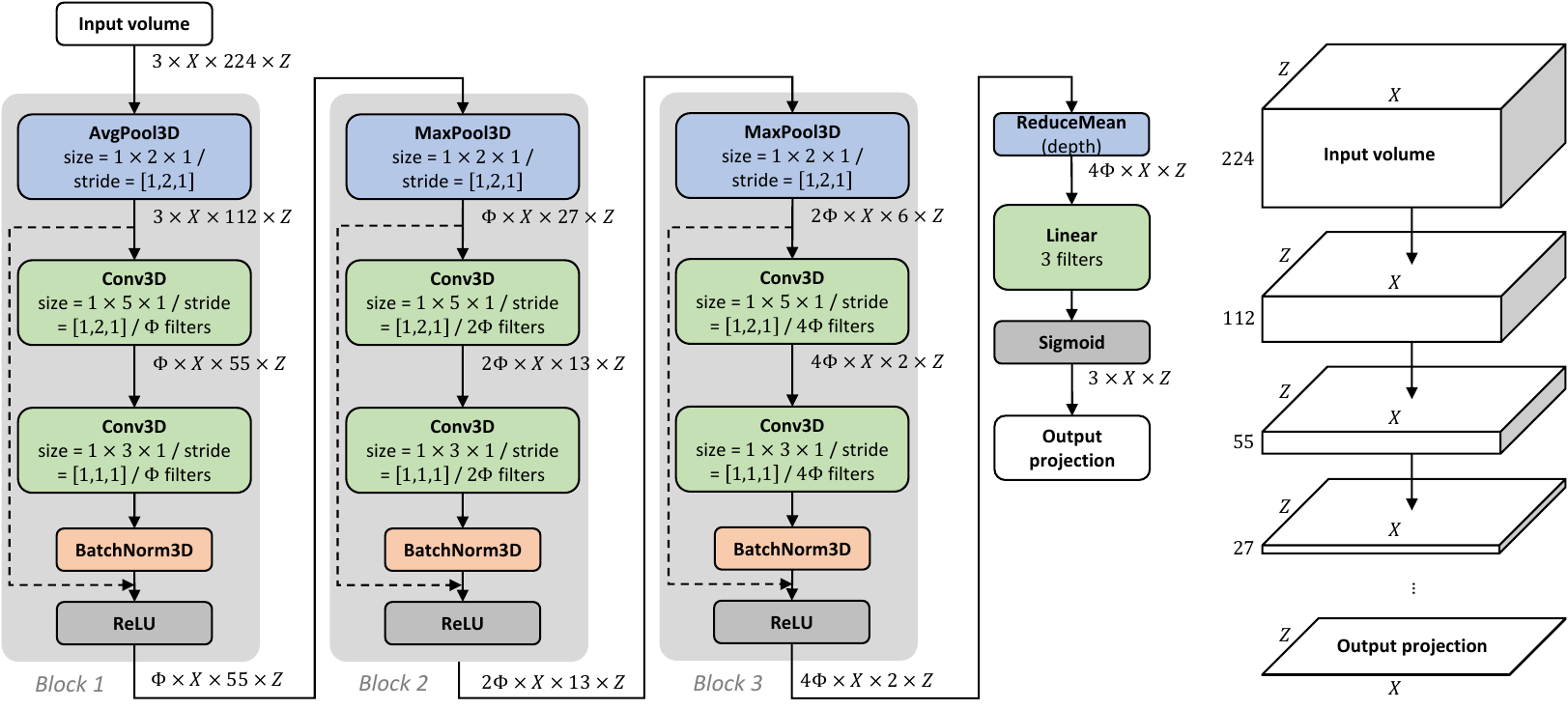}
    \caption{Architecture of the 3-D $\rightarrow$ 2-D projection network, detailed in section \ref{sec:projection}. The input is the preprocessed acquisition, a 3-channel volume of $X \times Y_1 \times Z$ voxels, with $Y_1 = 224$ in this example. Parameter $\Phi$, the number of filters in the first block, controls the complexity of the projection network. The figure on the right illustrates the size of the data tensors at the output of each block.}
    \label{fig:projector}
\end{figure*}

The preprocessed 3-D acquisition $I$ is then converted to a 2-D summary image $\Pi(I)$ through a parametric \tdtd en-face projection $\Pi$. \citet{lachinov_projective_2021} proposed a U-Net-like architecture for $\Pi$, where the encoder part contains 3-D operations and the decoder part contains 2-D operations. U-Net-like architectures have smaller and smaller activation maps as we go deeper into the encoder part, and their size increases as we go deeper into the decoder part to finally reach the size of the input image. This contraction aims to increase the receptive field of deep encoder filters to better consider the context without increasing their size and, therefore, the number of network parameters. The drawback of this contraction is that small details are lost in the process. To recover those details, skip-connections are therefore introduced between encoder and decoder layers \citep{ronneberger_u-net:_2015}. However, this trick assumes that the ground truth signal contains small details. In particular, it requires a dense supervision signal. For a classification task, the class labels are the only supervision signals available for training $\Pi$.

Our solution to the problem is to ensure that the details in the en-face plane are never lost throughout the \tdtd projection process. In particular, we guarantee that the activation maps all have the same size in the en-face plane ($X\times Z$ pixels). Only the depth of these activation maps decreases as we go deeper in $\Pi$, to reach a final depth of 1 voxel (i.e., a 2-D image). To further prevent the loss of details in the en-face plane, we also limit the receptive field of the filters to one pixel in that plane. A simple solution based on 1-D operations, where each A-scan is processed independently, is presented hereafter. It should be noted that \citet{li_image_2020}'s projection network for segmentation also ensures that activation maps all have the same size in the en-face plane. However, \citet{li_image_2020} use 3-D convolution operators (kernel size: $3\times 3 \times 3$ voxels): with a receptive field larger than one voxel in that plane, there is no guarantee that details are preserved without a dense supervision signal.

The proposed network, illustrated in Fig. \ref{fig:projector}, is divided into basic blocks containing:
\begin{enumerate}
    \item a pooling operator,
    \item two convolutional layers,
    \item a batch normalization operator \citep{ioffe_batch_2015},
    \item an optional skip-connection \citep{he_deep_2016},
    \item a ReLU activation.
\end{enumerate}
Note that the pooling operator precedes the convolutional layers in order to limit network complexity: since no contraction in the en-face plane is performed, this is critical. An average pooling operator is used in the first block; otherwise, half of the voxels would never be used. However, a max pooling operator is used in the following blocks to add more nonlinearity. Following common practice \citep{he_deep_2016}, the number of convolutional filters increases as we go deeper into the network. Let $\Phi$ denote the number of filters per layer in the first block. The number of filters per layer in the $i$-th block is set to $2^{i-1}\Phi$.

Each block reduces the depth by a factor of 4 (2 due to pooling $\times$ 2 due to the stride in the first convolution layer). After three blocks, a global mean operator along the depth axis is performed to eliminate the depth dimension. Finally, a dense layer with sigmoid activation is applied to obtain a 2-D image with the desired number of channels, namely three channels (see section \ref{sec:overview}). The sigmoid activation facilitates conversion to a bitmap image for visualization.

\subsection{Classification of the \tdtd Projection}
\label{sec:classification_projection}

\begin{figure}[!t]
    \centering
    \includegraphics[width=.475\textwidth]{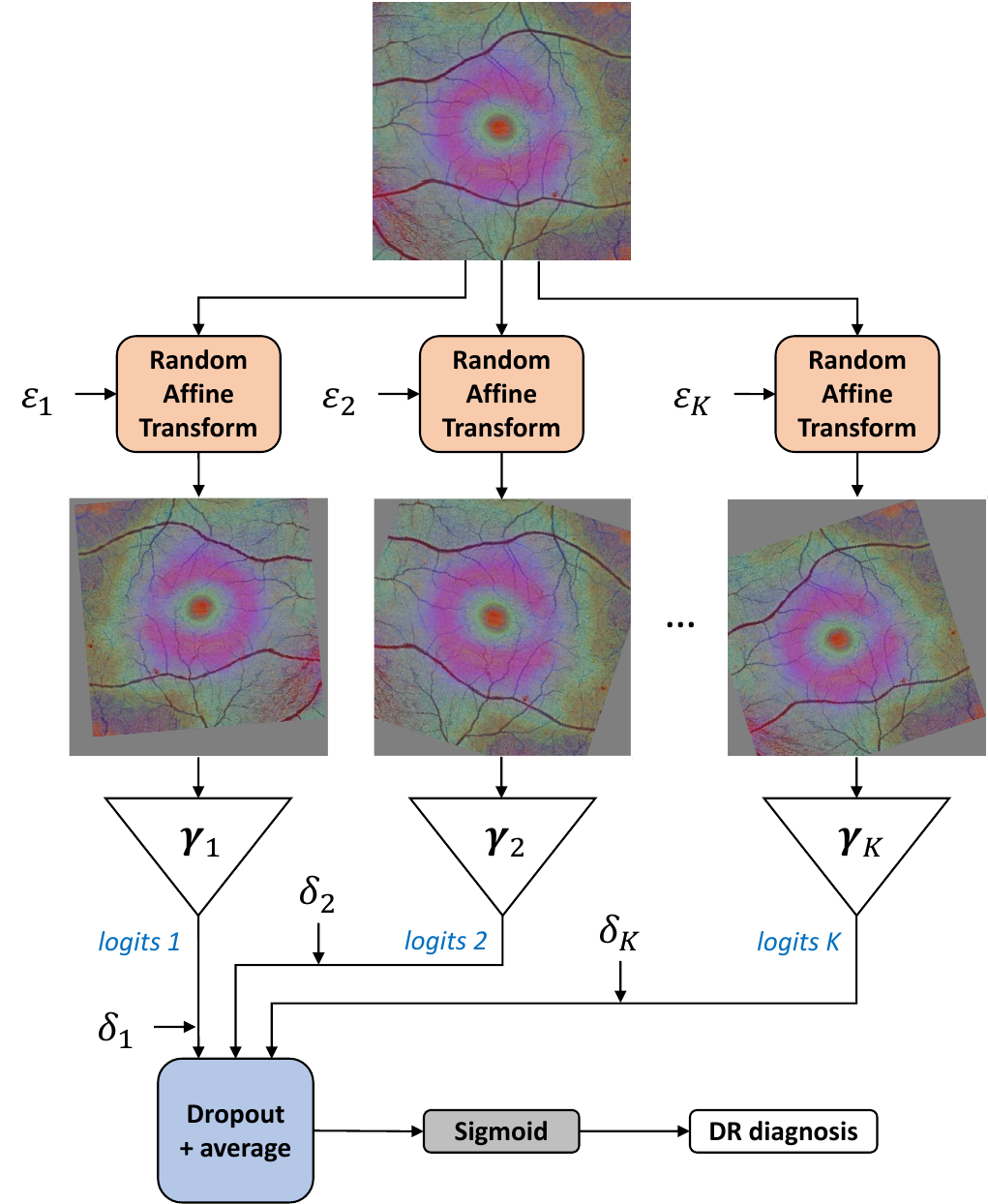}
    \caption{Ensemble of classification networks $\left\lbrace \boldsymbol{\gamma}_k, k = 1..K\right\rbrace$ with model dropout (controlled by random parameters $\delta_k, k = 1..K$) and differentiable random transformations (affine transformations and horizontal flips, controlled by random parameters $\varepsilon_k, k = 1..K$). This pipeline, detailed in section \ref{sec:classification_projection}, is illustrated for the first classification branch of Fig. \ref{fig:overview}, in which the input images are 2-D summary images.}
    \label{fig:ensemble}
\end{figure}

Now that a 2-D color image $\Pi(I)$ is obtained, any image classifier $C_1$ can be used to predict DR severity: a CNN, a transformer, an ensemble of CNNs and/or transformers, etc. However, two novel contributions specifically suited to this proposed framework are presented hereafter and illustrated in Fig. \ref{fig:ensemble}.

\subsubsection{Ensuring \tdtd Projection Generality}
\label{sec:projection_generality}

For interpretation purposes, we would like $\Pi(I)$ to be as independent from the classifier as possible. The rationale is as follows: if the projection is useful for any classifier, then we expect it to be informative for human experts as well.
Various solutions can be considered:
\begin{itemize}
  \item Following federated learning, multiple $\left\lbrace \Pi^{(j)}, C_1^{(j)}\right\rbrace$ couples can be trained in parallel, and the weights of the $\Pi^{(j)}$ instances can be aggregated at regular intervals.
  \item Following continual learning, multiple classifiers $C_1^{(j)}$ can be trained sequentially, initializing training with the weights obtained for $\Pi$ with the previous $C_1^{(j-1)}$ classifier.
\end{itemize}
However, such approaches imply longer training or require more resources. Instead, we propose to train one ensemble of classifiers, but with one trick that we call \textit{model dropout}: for each mini-batch, a random subset of the classifiers is used for prediction. Like the other solutions mentioned above, this ensures that the \tdtd projection $\Pi$ does not specialize for one specific classifier or one static ensemble of classifiers. Let $\boldsymbol{\gamma}_k, k = 1..K,$ denote the classifiers of the ensemble. We assume that these classifiers have no final activation function (i.e., they return logits). The ensemble prediction is given by:
\begin{align}
    \label{eq:ensemble}
    C_1\circ \Pi(I) & = \sigma\left( \frac{ \sum_{k=1}^K{ \delta_k \boldsymbol{\gamma}_k(\Pi(I)) } }{ \sum_{k=1}^K{ \delta_k } } \right) \\
    \text{subject to:}
        &\ \delta_k \in \{ 0, 1 \}, k=1..K \;\; , \notag \\
        & \textstyle 1 \leq \sum_{k=1}^K{ \delta_k } \leq K \;\; , \notag
\end{align}
where $\sigma$ denotes the sigmoid function. The number of possible classifier combinations is given by $2^K - 1$. This process is equivalent to training $2^K - 1$ classifiers in random order, which is expected to improve the generality of $\Pi$. Model dropout is only used during training: the full ensemble is used during inference. The expected benefit of logit averaging (see Eq. (\ref{eq:ensemble})) is that the weight of poor classifiers in the ensemble can be reduced automatically by decreasing the amplitude of their logits. However, logit averaging may have drawbacks \citep{tassi_impact_2022}, so the traditional strategy, namely probability averaging, was also investigated.

\subsubsection{Data Augmentation}
\label{sec:data_augmentation}

Data augmentation is typically performed by randomly transforming preprocessed images before feeding them to the neural network. For 2-D image classifiers, random transformations traditionally imply random affine transformations (random rotation, translation, and scaling) and random horizontal/vertical flips. However, our input preprocessed images are heavy 3-D volumes. Applying such random transformations to the 3-D volume takes a lot of time. Instead, we propose to apply them after the \tdtd projection: applied to 2-D data, they are much faster. Besides, applying random spatial transformations before the projection is of limited interest since the proposed projection operator $\Pi$ does not consider the context. Inserting random transformations inside the neural architecture is made possible by differentiable implementations of these transformations\footnote{\url{https://pytorch.org/vision/main/transforms.html}}.

Since these random transformations can be inserted inside the neural network, we are able to generate one transformed version of $\Pi(I)$ for each classifier $\boldsymbol{\gamma}_k$ in the ensemble. This leads to a new definition for $C_1\circ \Pi$:
\begin{align}
    \label{eq:classifier1}
    C_1\circ \Pi(I) & = \sigma\left( \frac{ \sum_{k=1}^K{ \delta_k \boldsymbol{\gamma}_k(T(\Pi(I), \varepsilon_k)) } }{ \sum_{k=1}^K{ \delta_k } } \right) \\
    \text{subject to:}
        &\ \delta_k \in \{ 0, 1 \}, k=1..K  \;\; , \notag \\
        & \textstyle 1 \leq \sum_{k=1}^K{ \delta_k } \leq K  \;\; , \notag
\end{align}
where $T$ denotes the transformation operator and $\varepsilon_k$ the random transformation parameters drawn for $\boldsymbol{\gamma}_k$. As a way to generalize test-time data augmentation \citep{krizhevsky_imagenet_2012}, random transformations are applied during both training and inference.

\subsection{Relevant B-scan Selection}
\label{sec:bscan_selection}

A first estimation $\mathbf{p}^{(1)}(I) = C_1 \circ \Pi(I)$ of the probabilistic prediction $\mathbf{p}(I)$, based on the en-face \tdtd projection $\Pi$, is now available. We propose to investigate further those B-scans of $I$ which contribute the most to $\mathbf{p}^{(1)}(I)$. The idea is to find additional evidence to increase or decrease the confidence in this first estimation.

To detect the B-scans that contribute the most to $\mathbf{p}^{(1)}(I)$, we propose to use attribution methods presented in section \ref{sec:attribution}. Note that attributions are computed for one particular output prediction $p_n^{(1)}(I)$, i.e., for one DR severity cutoff. This aligns with our goal to collect additional evidence for each prediction: we will select one B-scan per prediction. As for the inputs, we can either:
\begin{itemize}
    \item use the 3-D preprocessed acquisition $I$ and accumulate voxel-wise attributions in the xy-plane.
    \item or use the 2-D projection $\Pi(I)$ and accumulate pixel-wise attribution along the x-axis (the fast scanning axis).
\end{itemize}
The second option was chosen for faster computations. Let $a_I(x, z, c, n)$ denote the attribution of pixel $(x, z)$, in the $c$-th channel, for the $n$-th prediction. A normalized attribution $\alpha_I(z, n)$ is defined for the $z$-th B-scan, with respect to the $n$-th prediction:
\begin{equation}
    \alpha_I(z, n) = \frac{ \sum_x{\sum_c{| a_I(x, z, c, n) |}} }{ \sum_z{ \sum_x{\sum_c{| a_I(x, z, c, n) |}} } } \;\; .
    \label{eq:a-scan-attribution}
\end{equation}
Let $z_n$ denote the index of the $n$-th selected B-scan and let $B_n(I)$ denote its content:
\begin{equation}
    B_n(I) = I(c, x, y, z_n), \forall c, x, y  \;\; .
\end{equation}
For inference, the B-scans maximizing $\alpha_I(z, n)$, $n=1..N$, are selected. However, for data augmentation purposes and to favor exploration, a random B-scan selection process is preferred during training: $z_n$ is randomly drawn from the multinomial probability distribution defined by $\alpha_I(z, n)$:
\begin{align}
    \label{eq:selection}
    z_n & = \argmax_{z}{\alpha_I(z, n)} & \text{ for inference}  \;\; , \\
    z_n & \sim M_Z(1; \alpha_I(1, n), ..., \alpha_I(Z, n)) & \text{ for training}  \;\; .
\end{align}
It should be noted that B-scan selection is not impacted by random transformations from section \ref{sec:data_augmentation}, which are an integral part of the $C_1$ classifier on which the attribution method operates (see Eq. (\ref{eq:classifier1})).

\subsection{Final Classification}
\label{sec:second_final_classification}

\subsubsection{Second Classifier}
\label{sec:second_classification}
Like classifier $C_1$, the second classifier $C_2$ also requires data augmentation. Besides the random selection process described above, we propose to apply the same random transformation $T$ as for classifier $C_1$. More generally, we define $C_2$ very similarly to $C_1$: an ensemble of classifiers $\boldsymbol{\gamma}'_k$, $k=1..K$, with random transformations (parameters: $\varepsilon'_{k,n}$, $k=1..K$, $n=1..N$) and model dropout (parameters $\delta'_k$, $k=1..K$). Because their input images are of a different nature, no parameter sharing was set up between $C_1$ and $C_2$.

By design, the $n$-th selected B-scan $B_n(I)$ is meant to correct the confidence in the $n$-th prediction. Therefore, we only consider the $n$-th prediction $\gamma'_{k, n}(B_n(I))$ of classifier $\boldsymbol{\gamma}'$ for B-scan $B_n(I)$. This leads to the following expression for the predictions $\mathbf{p}^{(2)}(I)$ of $C_2$:
\begin{align}
    p^{(2)}_n(I) & = \sigma\left( \frac{ \sum_{k=1}^K{ \delta'_k \gamma'_{k, n}(T( B_n(I), \varepsilon'_{k,n} )) } }{ \sum_{k=1}^K{ \delta'_k } } \right), n=1..N \\
    \text{subject to:}
    &\ \delta'_k \in \{ 0, 1 \}, k=1..K \;\; , \notag \\
    & \textstyle 1 \leq \sum_{k=1}^K{ \delta'_k } \leq K \;\; . \notag
\end{align}

\subsubsection{Final Classifier}
\label{sec:final_classification}
The second classifier $C_2$ is supposed to increase or decrease the confidence in the predictions of the first classifier $C_1$. Therefore, the logits from both classifiers are combined linearly to obtain the final probabilistic prediction:
\begin{equation}
    \mathbf{p}(I) = \sigma\left( \sigma^{-1}( \mathbf{p}^{(1)}\left( I \right) ) + \sigma^{-1}( \mathbf{p}^{(2)}\left( I \right) ) \right) \;\; ,
\end{equation}
where $\sigma^{-1}$ is the logit function.

\subsubsection{Training}
\label{sec:training}
The multi-label classifier thus defined is trained to minimize the binary cross-entropy $\mathcal{L}$ between network predictions $p_n(I)$ and ground truth labels $\lambda_n(I)$, $n = 1..N$:
\begin{equation}
  \label{eq:crossEntropy}
  \begin{array}{rl}
    \mathcal{L} = -\frac{\displaystyle 1}{ N \displaystyle \sum_I{1} }
      \displaystyle\sum_I\sum_{n=1}^N & \left[\lambda_n(I)\log(p_n(I)) + \right. \\
                                      & \left. (1-\lambda_n(I))\log(1 - p_n(I)) \right] \;.
  \end{array}
\end{equation}
Thanks to this loss function, the ordered nature of DR severity grades is taken into account (see section \ref{sec:overview}). If a prediction is wrong by one severity level, then only one binary classifier will have an incorrect prediction. Whereas if a prediction is wrong by $n > 1$ severity levels, then $n$ binary classifiers will have an incorrect prediction, and will therefore impact the global loss $\mathcal{L}$ more.

We hypothesize that B-scan selection is most relevant when the first classifier $C_1$ is already well trained. Therefore, two training scenarios are investigated:
\begin{description}
    \item[Two-step training:] $C_1$ is trained alone until convergence. Then, its parameters are frozen, and $C_2$ is trained until convergence.
    \item[One-step training:] $C_1$ and $C_2$ are trained jointly until convergence.
\end{description}

This concludes the presentation of the proposed framework.

\section{Experiments}
\label{sec:experiments}

This framework is now evaluated for the task of automated DR severity assessment, according to the ICDR scale \citep{wilkinson_proposed_2003}, using OCTA.

\subsection{Dataset}
\label{sec:dataset}
In this study, we used OCTA images from the "Évaluation Intelligente de la Rétinopathie diabétique" (EviRed) project\footnote{\url{https://evired.org/}}, which comprises data collected between 2018 and 2022 from 14 hospitals and recruitment centers in France. \\
The PLEX Elite 9000, with a scanning frequency of 200 kHz, was employed to capture Swept-Source (SS) OCTA images. The ocular data in the EviRed dataset typically include two acquisition types: high-resolution 6x3x6 mm$^3$ SS-OCTA (500x1536x500 voxels) centered on the macula and lower-resolution 15x6x15 mm$^3$ ultra-widefield UWF-SS-OCTA (834x3072x834 voxels). Each OCTA volume contains 2-D en-face localizer, structural, and flow information. The EviRed dataset encompasses 811 eyes from 477 patients and is divided into training, validation, and test sets. It should be noted that, for a few eyes, we have only 6x6 mm$^2$ or 15x15 mm$^2$ acquisitions. The distribution of patients and images in each set is presented in Table \ref{tab:dataset_distribution}.

The partitioning of the EviRed dataset into training, validation, and test sets followed a systematic approach to ensure a robust evaluation of the models. The process was guided by the following criteria:

\begin{enumerate}
    \item Patient independence: To minimize the risk of data leakage, each patient's data was assigned to only one of the sets (training, validation, or test). This approach prevents the model from learning any patient-specific characteristics that could lead to overfitting or an inflated performance metric.
    \item Balanced distribution of disease severity: The dataset was divided in such a way that each set contained a similar proportion of images representing various stages of DR. This balanced distribution ensures that the model is exposed to a wide range of severity levels during the training process and provides a more accurate assessment of its performance during validation and testing.
    \item Stratified sampling: To maintain consistency in the demographic characteristics and other factors, stratified sampling was employed when dividing the dataset. This approach not only ensures that each set contains a representative sample of the entire dataset but also respects the original class distribution in each split. By mirroring the class distribution of the whole dataset within each subset, we further reduce the risk of biased performance evaluation. Hence, we achieve a balanced representation of demographic characteristics and other factors across all subsets. This method gives us the confidence that the inferences drawn from our study will be robust and reliable.
\end{enumerate}

\begin{table}[!h]
\centering
\begin{tabular}{|l|ccc|ccc|}
\hline
 & \multicolumn{3}{c|}{PLEX Elite 6x6mm$^{2}$} & \multicolumn{3}{c|}{PLEX Elite 15x15mm$^{2}$} \\
\cline{2-7}
 & Patients & Eyes & \% & Patients & Eyes & \% \\
\hline
\hline
Train & 333 & 625 & 70.0 & 318 & 584 & 70.2 \\
Val & 82 & 159 & 17.7 & 78 & 147 & 17.7 \\
Test & 57 & 110 & 12.3 & 57 & 100 & 12.1 \\
\hline
\end{tabular}
\caption{Dataset distribution among training, validation, and test sets for PLEX Elite 6x6mm$^{2}$ and PLEX Elite 15x15mm$^{2}$}
\label{tab:dataset_distribution}
\end{table}

Because 6x6mm$^2$ and 15x15mm$^2$ acquisitions have different sizes and resolutions, distinct models were built for these two types of acquisitions.

\subsection{Implementation}
\label{sec:implementation}

The proposed framework was implemented using PyTorch\footnote{\url{https://pytorch.org/}} Ignite for training and inference, MONAI\footnote{\url{https://monai.io/}} for 3-D data handling, PyTorch Torchvision for differentiable data augmentation, PyTorch Image Models (timm)\footnote{\url{https://github.com/fastai/timmdocs/}} for 2-D image classification and Captum\footnote{\url{https://captum.ai/}} for attribution methods. Experiments were performed using two NVIDIA V100 GPUs (with 32 Gb of RAM each). One of the GPUs was dedicated to \tdtd projection ($\Pi$), and the other one was dedicated to 2-D image classification ($C_1$ and $C_2$).

\subsection{Hyperparameter Optimization}
\label{sec:hyperopt}

Various hyperparameters need to be set:
\begin{itemize}
  \item Architecture parameters: the use of $C_1$ alone or the joint use of $C_1$ and $C_2$ (see section \ref{sec:overview}).
  \item Preprocessing parameters: the depth $Y_0$ at which the ILM is aligned and the depth $Y_1$ under which the volume is cropped (see section \ref{sec:preprocessing}).
  \item \tdtd projection parameters: the number $\Phi$ of filters per convolutional layer in the first block and the use, or not, of skip-connections within blocks (see section \ref{sec:projection}).
  \item Data augmentation parameters: the range of values for random affine transformations $\varepsilon$ and $\varepsilon'$ (see sections \ref{sec:classification_projection} and \ref{sec:second_classification}).
  \item Ensemble parameters: the list of $K$ classification backbones in $C_1$ and $C_2$, and the use of logit or probability averaging (see sections \ref{sec:classification_projection} and \ref{sec:second_classification}).
  \item B-scan selection parameters: the attribution method used (see section \ref{sec:bscan_selection}).
  \item Training parameters: the general training parameters (optimizer, learning rate, etc.), training from scratch or the use of ImageNet pre-trained weights, and the problem-specific training schedule (one-step or two-steps --- see section \ref{sec:training}).
\end{itemize}

The preprocessing and data augmentation parameters were set empirically through visual inspection: $Y_0 = 32$ voxels, $Y_1 = 224$ voxels, random rotation in the range $[-10; +10]$ degrees, random translation in the range $[-10; +10]$ percent of $X$ and $Y_1$ or $Z$, random scaling in the range $[90; 110]$ percent of $X$ and $Y_1$ or $Z$. The following parameters were restricted due to GPU memory limitations and computation time considerations:
\begin{itemize}
    \item The number $\Phi$ of filters was limited to $\Phi \leq 32$ for 6x6mm$^2$ acquisitions, $\Phi \leq 16$ for 15x15mm$^2$ acquisitions.
    \item Integrated Gradients and perturbation-based attribution methods could not be used during training. The following attribution methods were investigated: Saliency, Deconvolution, Guided Backprop, and DeepLIFT.
\end{itemize}
The classifier backbones were chosen among CNNs: one advantage of most CNNs is that they can be applied to images of any size (in our case: $500 \times 500$ or $834 \times 834$ pixels) without adaptation. Two ensembles of classifiers were considered, with the following pre-trained weights (unless training from scratch is experimented):
\begin{description}
    \item[Ensemble 1:] The first set of classifiers was chosen from a single family, namely EfficientNet \citep{tan_efficientnet_2019}. This family is popular for its good trade-off between complexity and performance. The $K=4$ simplest networks were selected: namely EfficientNet-\{B0, B1, B2, B3\}, pre-trained on ImageNet-1K. This first ensemble was intended for quick experiments, hyper-parameter optimization, etc.
    \item[Ensemble 2:] The second set of classifiers was chosen among the best-performing CNN families in ImageNet classification benchmarks.\footnote{\url{https://github.com/kentaroy47/timm_speed_benchmark} --- page accessed on January 2023.} The set includes ConvNeXt Base, pre-trained on ImageNet-21K \citep{liu_convnet_2022}, ImageNet-V2 Small, pre-trained on ImageNet-21K \citep{tan_efficientnetv2_2021} and SE-ResNet-152D, pre-trained on ImageNet-1K \citep{hu_squeeze-and-excitation_2020}. The set was limited to $K=3$ due to GPU memory limitations.
\end{description}

The values of the other hyperparameters were then optimized using ensemble 1 for 6x6mm$^2$ acquisitions. Optimization relied on a Receiver Operating Characteristic (ROC) analysis in the validation set: hyperparameter values maximizing the Area Under the ROC Curve (AUC), averaged over the $N=4$ binary classification tasks, were selected. The following hyperparameters were obtained:
\begin{itemize}
    \item Full architecture ($C_1$ and $C_2$).
    \item $\Phi = 8$ filters in the first block (and therefore $2\Phi = 16$ filters in the second and $2^2 \Phi = 32$ filters in the third), without skip-connections.
    \item Logit averaging.
    \item Guided Backprop attribution method.
    \item Adam optimizer with an initial learning rate of $10^{-3}$ and an exponential learning rate scheduler ($99\%$ multiplicative decay at every epoch), using pre-trained weights.
    \item Two-step training.
\end{itemize}

\subsection{Analysis of \tdtd Projections}

\begin{figure*}[!p]
    \centering
    \includegraphics[width=\textwidth]{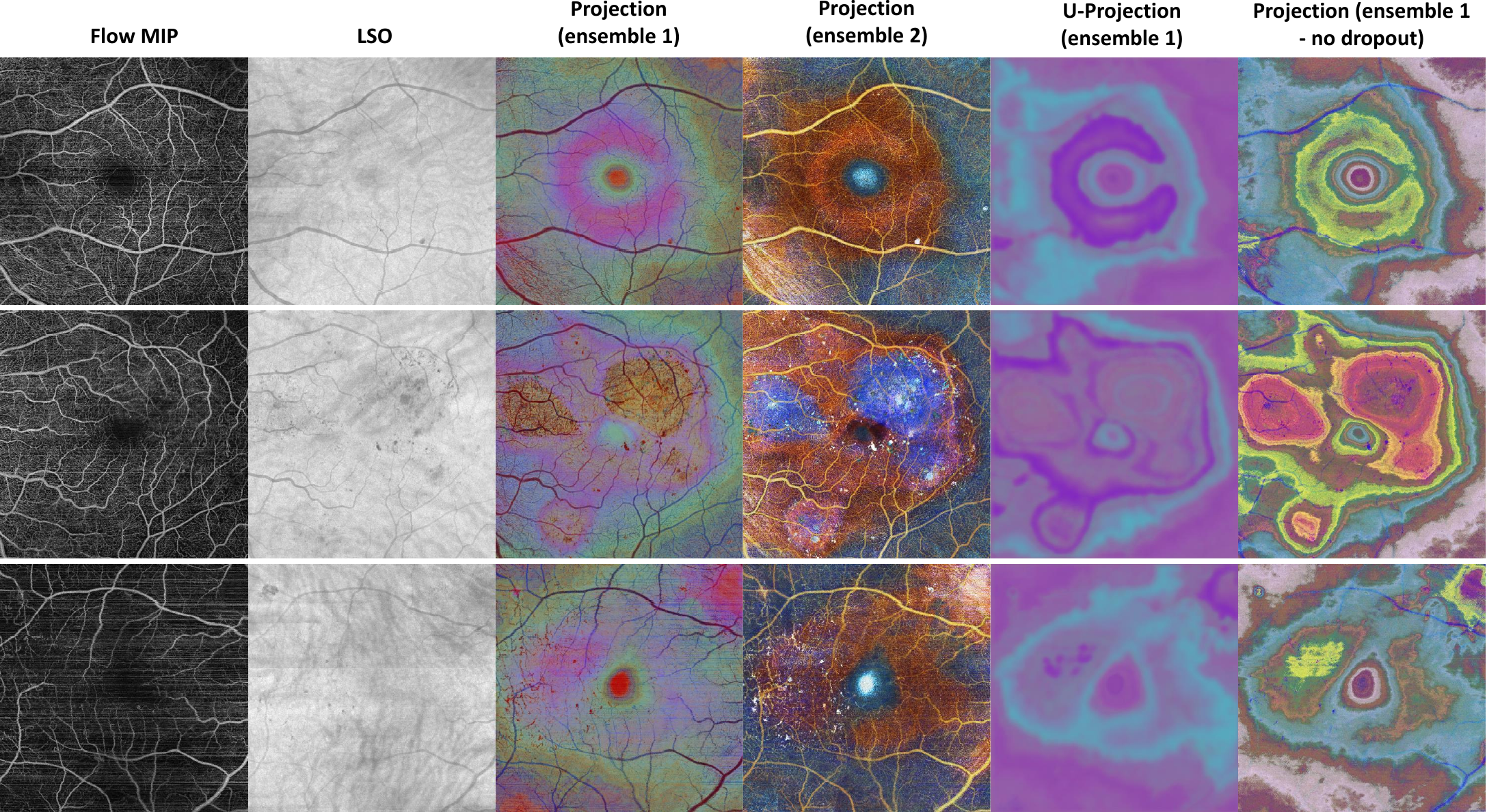}
    \caption{Examples of projections for 6x6mm$^2$ acquisitions from the test set. The projector has $\Phi=8$ filters in the first block and the classifiers are EfficientNet-\{B0, B1, B2, B3\} for ensemble 1 and \{ConvNeXt-base, Efficient-v2, SEResNet-152\} for ensemble 2. The first row represents an eye graded as moderate NPDR. The second row represents an eye graded as moderate NPDR with macular edema. The third row represents an eye graded as proliferative DR.}
    \label{fig:6x6_examples1}
\end{figure*}

\begin{figure*}[!p]
    \centering
    \includegraphics[width=\textwidth]{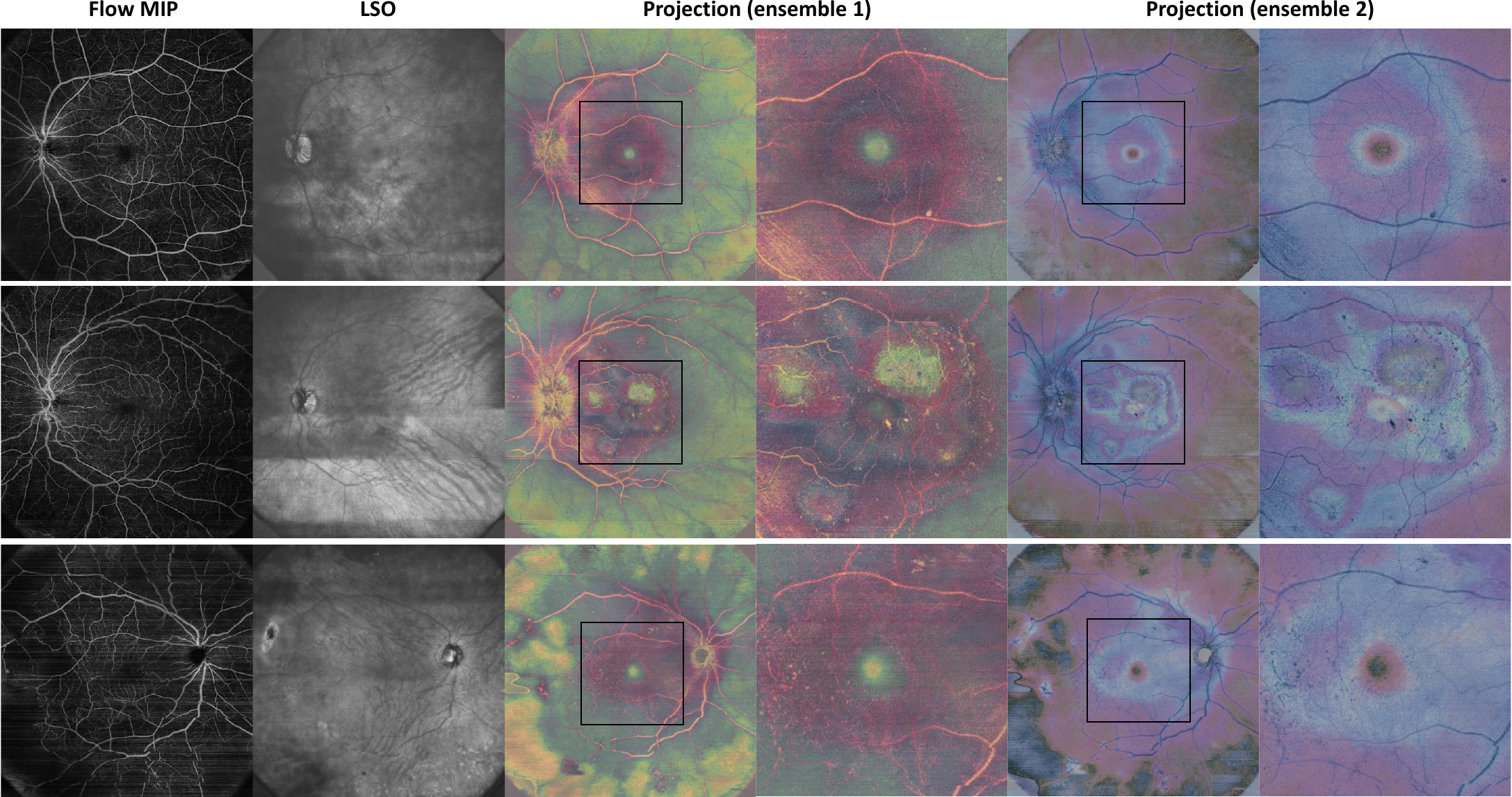}
    \caption{Examples of projections for 15x15mm$^2$ acquisitions using the same hyperparameters as in Fig. \ref{fig:6x6_examples1}, for the same eyes (the models are retrained for the new acquisition size). The squares indicate the zoomed areas, which were imaged in the 6x6mm$^2$ acquisitions.}
    \label{fig:15x15_examples1}
\end{figure*}

\begin{figure}[!t]
    \centering
    \includegraphics[width=0.4\textwidth]{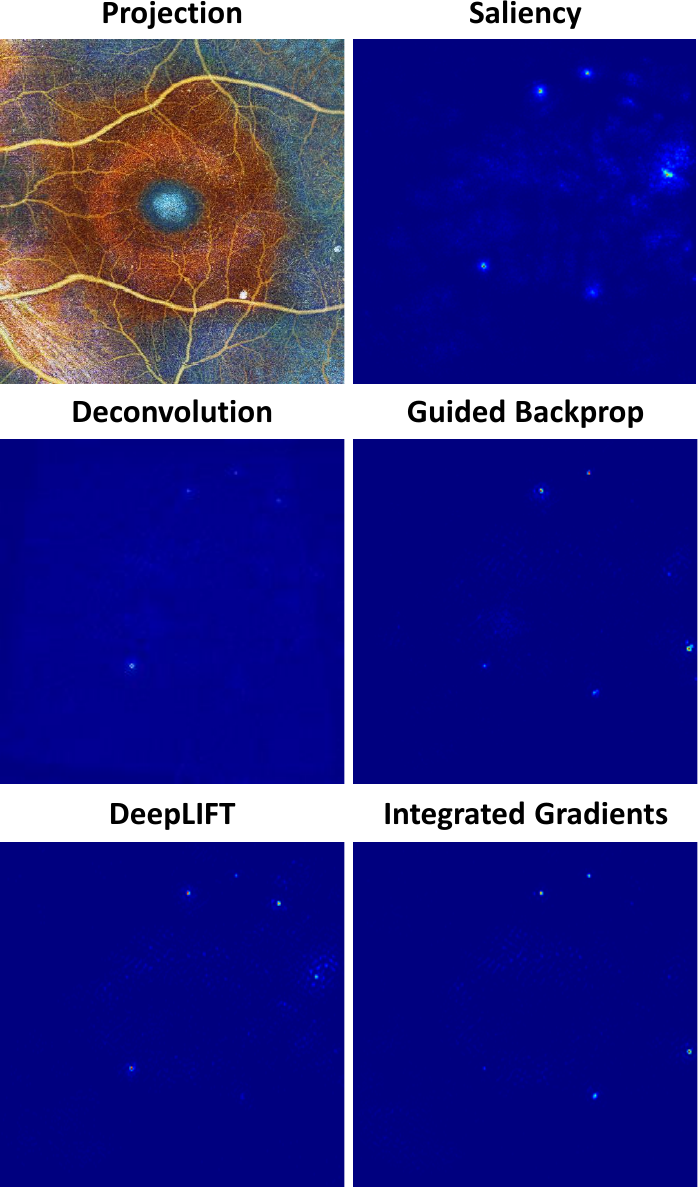}
    \caption{Comparison of a few attribution methods described in section \ref{sec:attribution}, computed for a 6x6mm$^2$ acquisition ($\Phi=8$, ensemble 2: \{ConvNeXt-base, Efficient-v2, SEResNet-152\}). Each pixel intensity represents the sum of attributions computed for the $N=4$ model outputs (one output per DR severity cutoff), see Eq. (\ref{eq:a-scan-attribution}). The `jet' color map, ranging from blue (0) to red (255), is used.}
    \label{fig:attribution}
\end{figure}

Examples of \tdtd projections obtained for 6x6mm$^2$ acquisitions using both ensembles of classifiers are presented in Fig. \ref{fig:6x6_examples1}. They are compared with projections obtained using the U-Net-like projection (or U-Projection) proposed by \citet{lachinov_projective_2021} for \tdtd segmentation, and also with projections obtained without model dropout. Examples of \tdtd projections obtained for 15x15mm$^2$ acquisitions of the same eyes are presented in Fig. \ref{fig:15x15_examples1}. Next, we present in Fig. \ref{fig:attribution} examples of A-scan-level attributions (see Eq. (\ref{eq:a-scan-attribution})) obtained for $C_1$ with various attribution methods. Attributions were obtained for 6x6mm$^2$ acquisitions using ensemble 2. This figure illustrates that the Guided Backprop, DeepLIFT, and Integrated Gradients methods produce rather similar results. The Saliency and Deconvolution methods produce more noisy results.

\begin{figure*}[!h]
    \centering
    \includegraphics[width=0.7\textwidth]{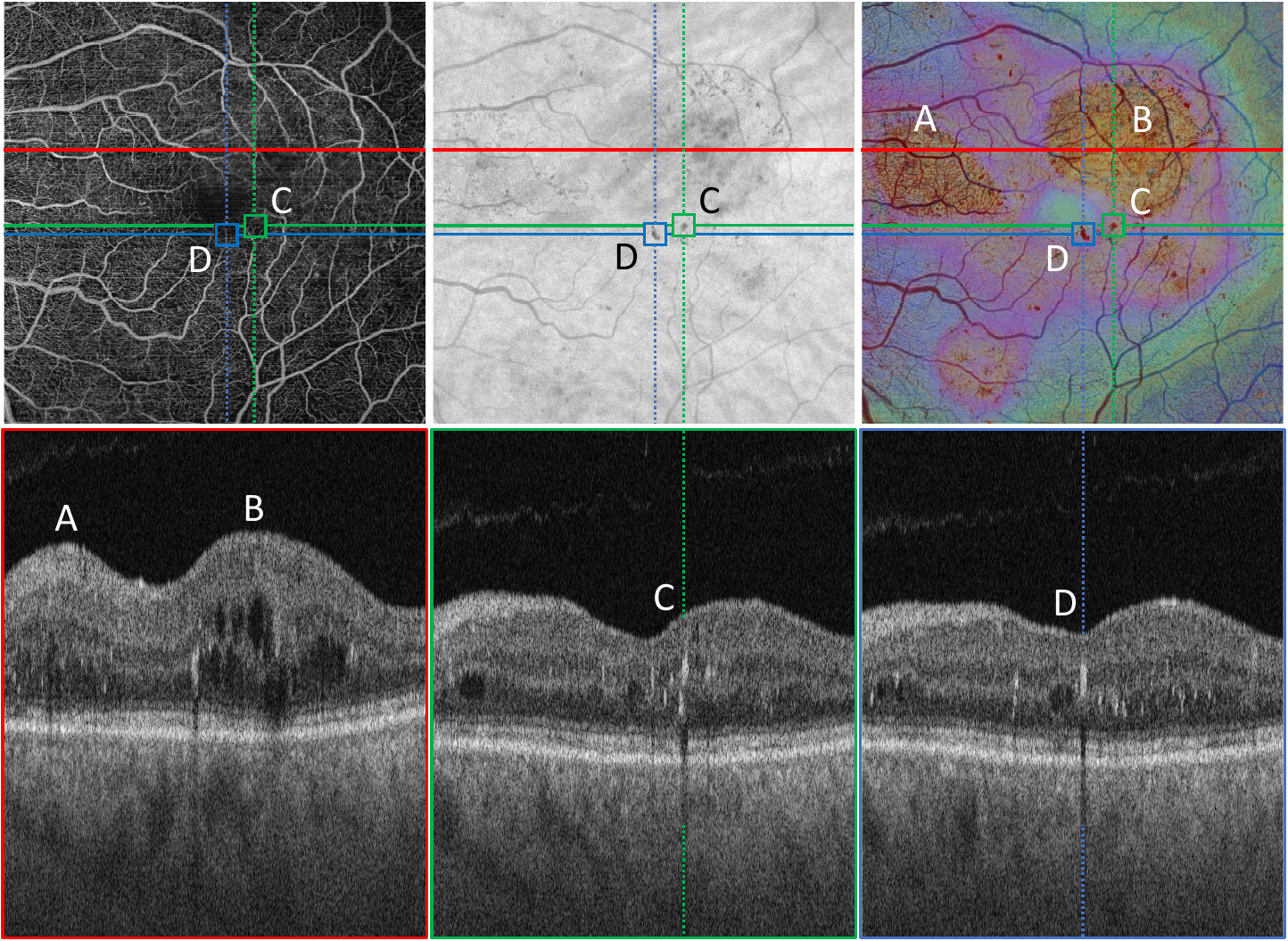}
    \caption{Retinal lesions highlighted in the 2-D projections (for the second eye of Fig. \ref{fig:6x6_examples1} and \ref{fig:15x15_examples1}). The first row shows, from left to right, the flow MIP, the LSO, and the projection (ensemble 1). The second row shows B-scans. Macular edemas (A and B) are visible in the first B-scan (horizontal red line). Large microaneurysms (C and D) are visible in the following B-scans (horizontal green and blue lines, respectively).}
    \label{fig:evidence}
\end{figure*}

\subsection{Performance evaluation}
\label{sec:performance_evaluation}

\begin{figure*}[!t]
    \centering
    \begin{tabular}{cc}
        \includegraphics[width=.375\textwidth]{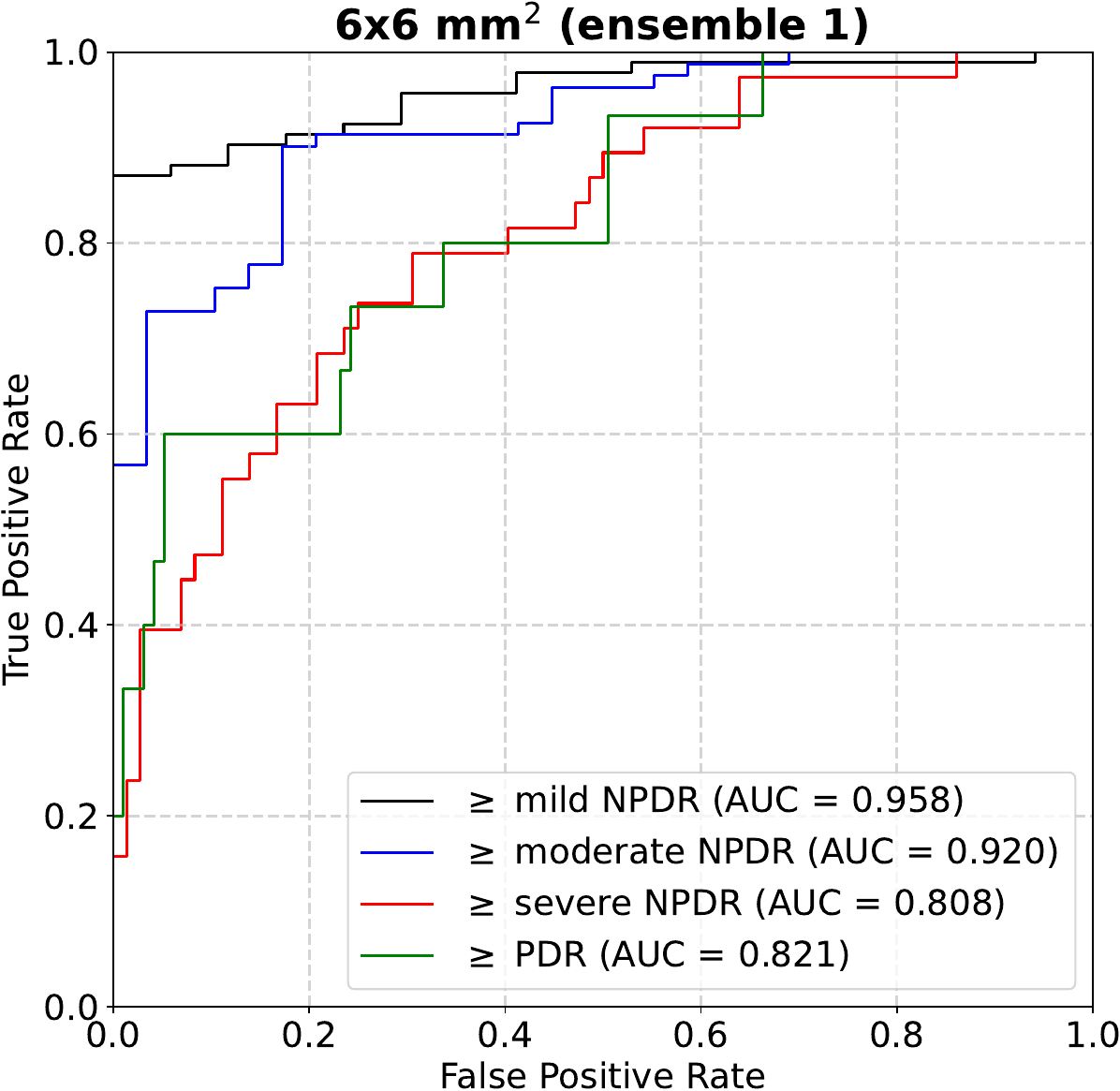} & \includegraphics[width=.375\textwidth]{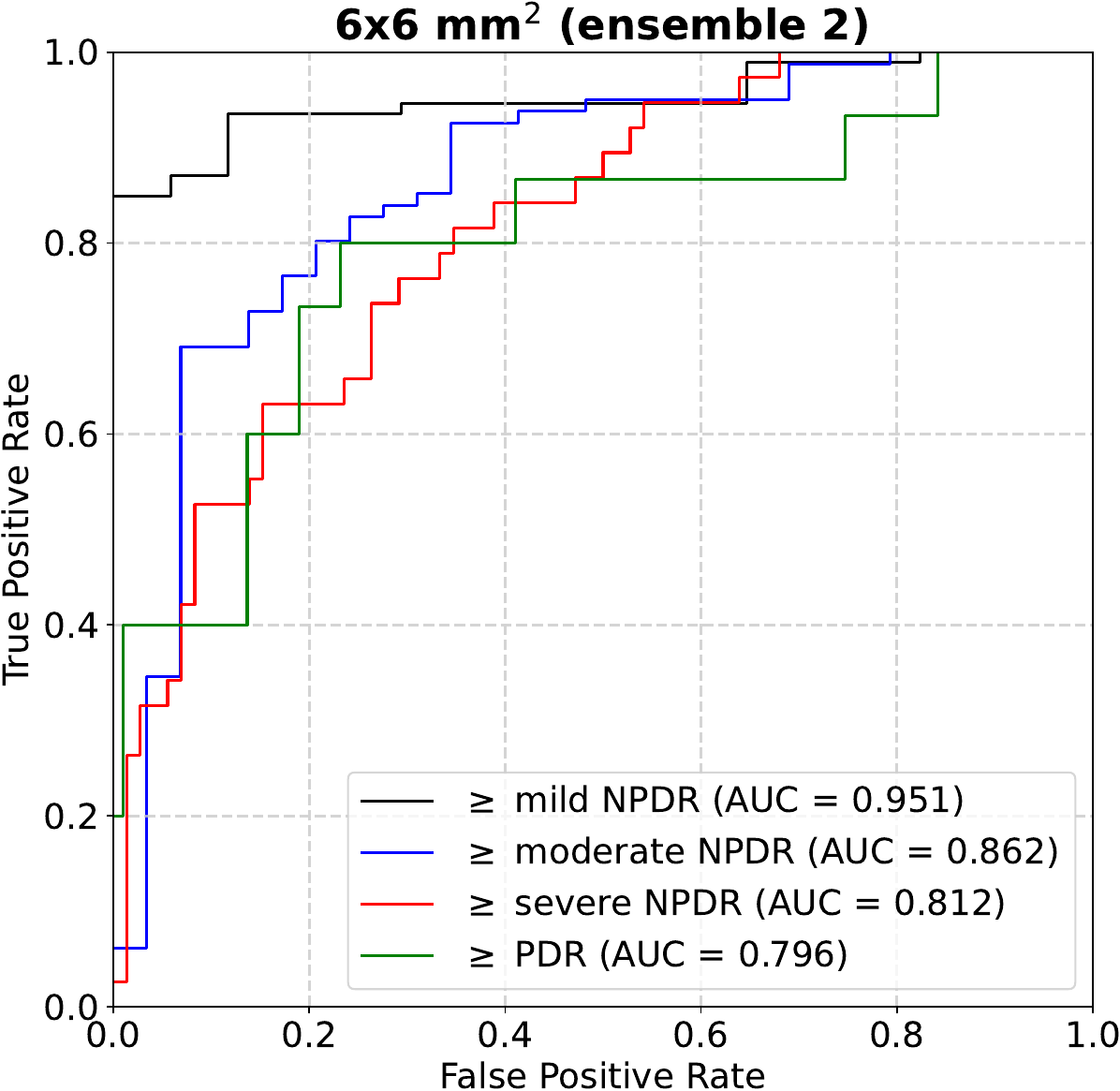} \\
        \includegraphics[width=.375\textwidth]{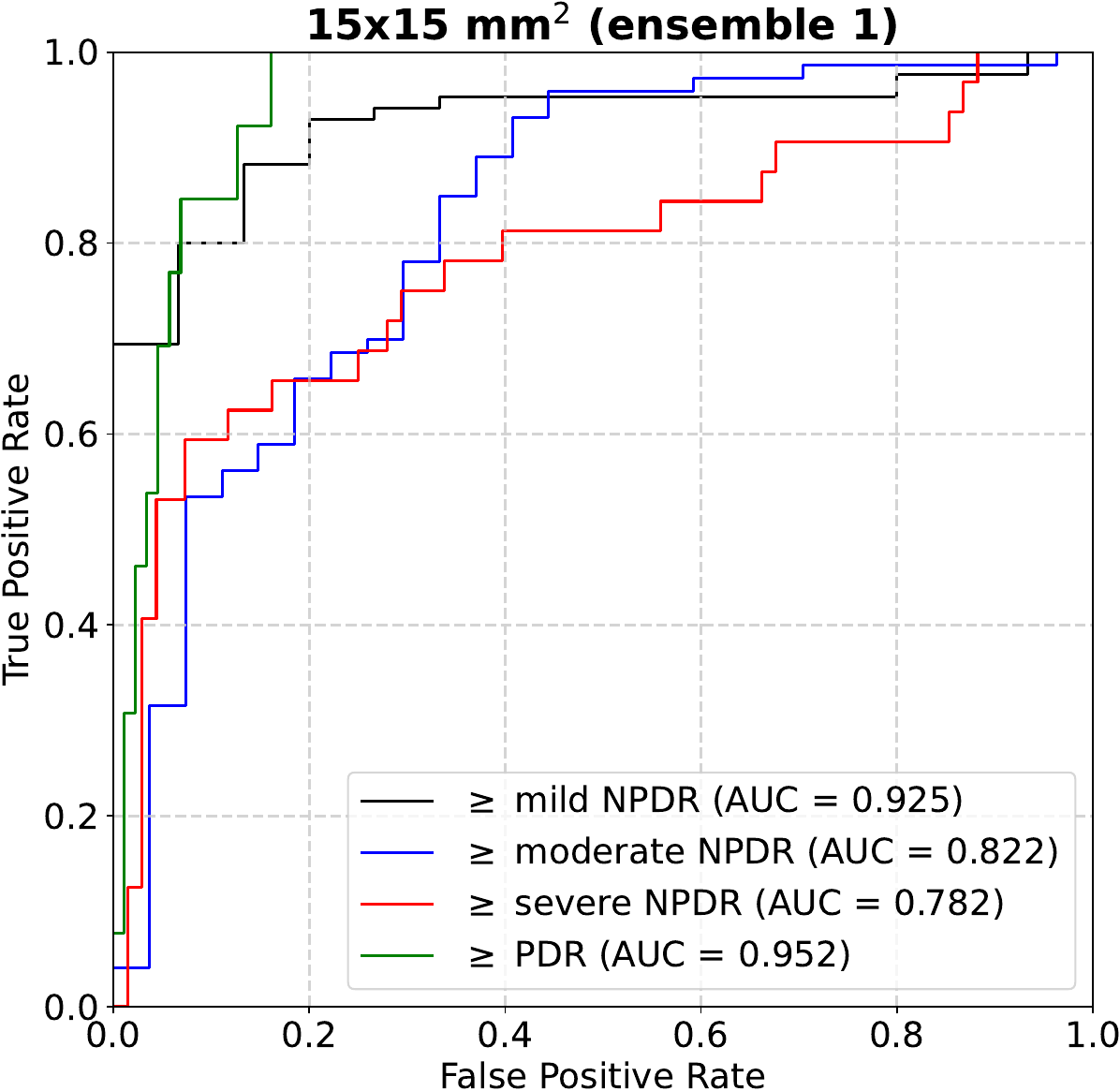} & \includegraphics[width=.375\textwidth]{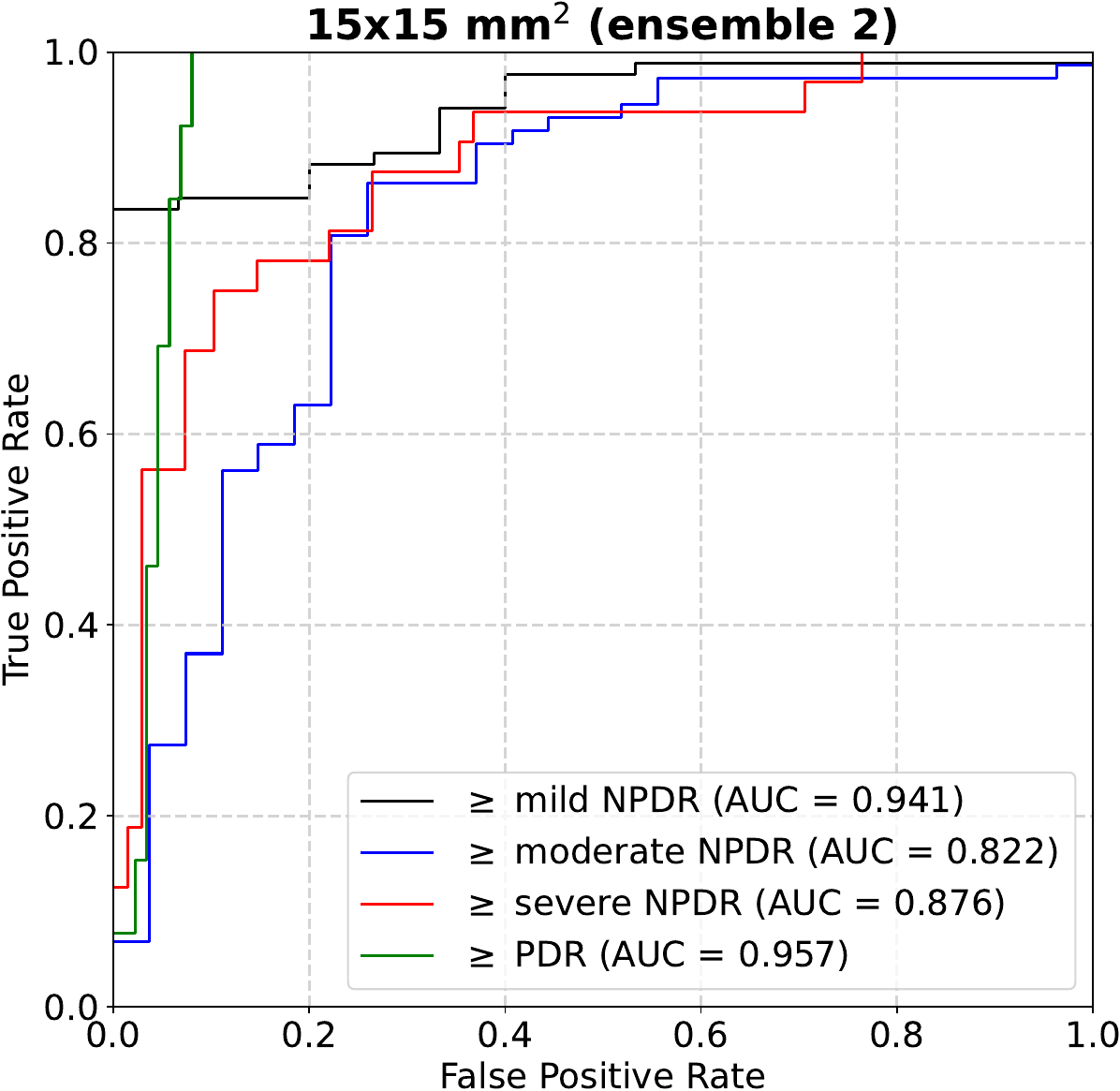} \\
    \end{tabular}
    \caption{Receiver Operating Characteristic of the proposed system in the test set using EfficientNet-\{B0, B1, B2, B3\} (ensemble 1) or \{ConvNeXt-base, Efficient-v2, SEResNet-152\} (ensemble 2) for 6$\times$6mm$^2$ and 15$\times$15mm$^2$ acquisitions.}
    \label{fig:roc}
\end{figure*}

Classification performance achieved in the test set for 6x6mm$^2$ and 15x15mm$^2$ acquisitions is presented in Table \ref{tab:performance} (a). Results are reported for both ensembles and all $N=4$ classification tasks. For a given ensemble and a given classification task, three AUC values are reported: the one obtained for the full classifier (predictions $\mathbf{p}(I)$) and the ones obtained for each branch $C_1$ and $C_2$ of the classifier separately (predictions $\mathbf{p}^{(1)}(I)$ and $\mathbf{p}^{(2)}(I)$, respectively). ROC curves for the full classifiers are reported in Fig. \ref{fig:roc}.

To show the relevance of the second branch (classifier $C_2$), we compared the AUC values obtained using 1) $C_1$ alone, 2) $C_2$ alone, or 3) $C_1$ and $C_2$ jointly. For each of these 3 scenarios, a set of 16 AUC values is available in Table \ref{tab:performance} (a)(2 acquisitions $\times$ 2 CNN ensembles $\times$ 4 decisions). Wilcoxon signed-rank tests were performed to compare two scenarios by confronting the corresponding 16 pairs of AUC values: results are reported in Table \ref{tab:performance} (b).

\begin{table*}[!t]

    \begin{tabular}{cc}

    \subfloat[Performance]{
    \centering
    \begin{tabular}{|c|c|c|cccc|}
        \hline
        Acquisition                         & Backbones          & Classification &  $\geq$ mild   & $\geq$ moderate & $\geq$ severe  & $\geq$ PDR     \\
                                            &                    & branches       &  NPDR          & NPDR            & NPDR           &                \\
        \hline
        \hline
        \multirow{6}{*}{6$\times$6mm$^2$}   & EfficientNet-      & $C_1$          & 0.946          & 0.913           & 0.809          & 0.815          \\
                                            & \{B0, B1, B2, B3\} & $C_2$          & 0.798          & 0.764           & 0.675          & 0.589          \\
                                            &                    & $\{C_1 ,C_2\}$ & \textbf{0.958} & \textbf{0.920}  & 0.808          & 0.821          \\
        \cline{2-7}
                                            & \{ConvNeXt-base,   & $C_1$          & 0.935          & 0.849           & 0.768          & 0.749          \\
                                            &   Efficient-v2,    & $C_2$          & 0.824          & 0.737           & 0.734          & 0.816          \\
                                            &   SEResNet-152\}   & $\{C_1 ,C_2\}$ & 0.951          & 0.862           & 0.812          & 0.796          \\
        \hline
        \hline
        \multirow{6}{*}{15$\times$15mm$^2$} & EfficientNet-      & $C_1$          & 0.918          & 0.815           & 0.767          & 0.948          \\
                                            & \{B0, B1, B2, B3\} & $C_2$          & 0.766          & 0.726           & 0.725          & 0.848          \\
                                            &                    & $\{C_1 ,C_2\}$ & 0.925          & 0.822           & 0.782          & 0.952          \\
        \cline{2-7}
                                            & \{ConvNeXt-base,   & $C_1$          & 0.912          & 0.800           & \textbf{0.880} & 0.947          \\
                                            &   Efficient-v2,    & $C_2$          & 0.867          & 0.774           & 0.658          & 0.676          \\
                                            &   SEResNet-152\}   & $\{C_1 ,C_2\}$ & 0.941          & 0.822           & 0.876          & \textbf{0.957} \\
        \hline
    \end{tabular}}

    \subfloat[Comparison]{
    \centering
    \begin{tabular}{|c|c||c|}
        \hline
        \multicolumn{2}{|c||}{Classification} & Wilcoxon \\
        \multicolumn{2}{|c||}{Branches}       & test ($p$) \\
        \hline
        \hline
        $C_1$ & $C_2$                         & 0.00031$^\ast$ \\
        \hline
        $C_1$ & $\{C_1 ,C_2\}$                & 0.00015$^\ast$ \\
        \hline
    \end{tabular}}

    \end{tabular}

    \caption{Performance of the proposed framework in terms of area under the ROC curve (AUC) in the test set (a). Impact of the classification branches on performance (b). $^\ast$Significant difference, according to a Wilcoxon signed-rank test on the paired AUC values ($p<0.05$).}
    \label{tab:performance}
\end{table*}

\subsection{Ablation Study}

Additional experiments were performed with 6x6mm$^2$ acquisitions, using ensemble 1, to study the impact of the main hyperparameters. Results of experiments investigating the first classification branch $C_1$ only are reported in Table \ref{tab:ablation_study_1}, and those investigating the second branch are reported in Table \ref{tab:ablation_study_2}. Please note that results are reported on the test, not on the validation set (used for hyperparameter optimization), so the ranking of solutions may contradict the optimal hyperparameter values listed in section \ref{sec:hyperopt} (associated with test results reported in Table \ref{tab:performance}).

In this ablation study, we only considered one acquisition and one CNN ensemble. With only four pairs of AUC values to compare, the previous statistical test is no longer suitable. Instead of comparing AUC values, we thus compared ROC curves directly: Delong tests were used for that purpose. Delong tests are designed for comparing two curves, not two sets of curves: each set of curves was thus micro-averaged \citep{aguilar-ruiz_multiclass_2022} prior to Delong testing.

\begin{table*}[!t]
    \centering
    \begin{tabular}{|cccccc|cccc|c|}
        \hline
        $\Phi$      & Number                 & U-           & Model       & From         & Probability  & $\geq$ mild & $\geq$ moderate & $\geq$ severe & $\geq$ PDR & Delong \\
                    & of filters             & Projection?  & dropout?    & scratch?     & averaging?   & NPDR        & NPDR            & NPDR          &            & test ($p$) \\
        \hline
        \hline
        8           & \{8,16,32\}            & no           & yes         & no           & no           & 0.946       & 0.913           & 0.809         & 0.815      & \\
        \hline
        \hline
        \textbf{4}  & \textbf{\{4,8,16\}}    & no           & yes         & no           & no           & 0.865       & 0.799           & 0.755         & 0.860      & 0.0401$^\ast$ \\
        \textbf{16} & \textbf{\{16,32,64\}}  & no           & yes         & no           & no           & 0.942       & 0.893           & 0.798         & 0.770      & 0.1416 \\
        \textbf{32} & \textbf{\{32,64,128\}} & no           & yes         & no           & no           & 0.937       & 0.913           & 0.836         & 0.839      & 0.9100 \\
        \hline
        8           & \{8,16,32\}            & \textbf{yes} & yes         & no           & no           & 0.851       & 0.774           & 0.754         & 0.769      & 0.0266$^\ast$ \\
        \hline
        8           & \{8,16,32\}            & no           & \textbf{no} & no           & no           & 0.856       & 0.834           & 0.718         & 0.731      & 0.0004$^\ast$ \\
        \hline
        8           & \{8,16,32\}           & no           & yes         & \textbf{yes} & no           & 0.804       & 0.752          & 0.725         & 0.854      & $<$0.0001$^\ast$ \\
        \hline
        8           & \{8,16,32\}           & no           & yes         & no           & \textbf{yes} & 0.921       & 0.867          & 0.836         & 0.817      & 0.0072$^\ast$ \\
        \hline
    \end{tabular}
    \caption{Influence of various hyperparameters on the area under the ROC curve (AUC) in the test set --- analysis of the first classification branch $C_1$ only (no B-scan selection). Experiments were performed for 6$\times$6mm$^2$ acquisitions and the EfficientNet-\{B0, B1, B2, B3\} backbones (ensemble 1). The first line corresponds to the optimal values (based on experiments on the validation set). Investigated parameters are in bold. U-Projection denotes the U-Net-like projection proposed by \citet{lachinov_projective_2021}. $^\ast$Significant difference with the reference (first line), according to a Delong test on the micro-averaged ROC curves ($p<0.05$).}
    \label{tab:ablation_study_1}
\end{table*}

\begin{table*}[!t]
    \centering
    \begin{tabular}{|ccc|cccc|c|}
        \hline
        One-step?    & B-scan          & Attribution            & $\geq$ mild  & $\geq$ moderate  & $\geq$ severe  & $\geq$ PDR & Delong \\
                     & selection       & method                 & NPDR         & NPDR             & NPDR           &            & test ($p$) \\
        \hline
        \hline
        no           & random          & Guided backprop        & 0.958        & 0.920            & 0.808          & 0.821      & \\
        \hline
        \hline
        \textbf{yes} & random          & Guided backprop        & 0.930        & 0.855            & 0.809          & 0.846      & 0.0004$^\ast$ \\
        \hline
        no           & \textbf{argmax} & Guided backprop        & 0.945        & 0.914            & 0.813          & 0.813      & 0.2411 \\
        \hline
        no           & random          & \textbf{Saliency}      & 0.954        & 0.920            & 0.808          & 0.815      & 0.8705 \\
        no           & random          & \textbf{Deconvolution} & 0.953        & 0.918            & 0.810          & 0.816      & 0.8541 \\
        no           & random          & \textbf{DeepLIFT}      & 0.955        & 0.918            & 0.811          & 0.825      & 0.9723 \\
        \hline
    \end{tabular}
    \caption{Influence of various hyperparameters on the AUC in the test set --- analysis of the second classification branch $C_2$ (jointly with $C_1$). Experiments were performed for 6$\times$6mm$^2$ acquisitions and the EfficientNet-\{B0, B1, B2, B3\} backbones (ensemble 1). The first line corresponds to the optimal values (based on experiments on the validation set). Investigated parameters are in bold. Argmax B-scan selection refers to a deterministic selection of B-scans during training, like during inference (see Eq. (\ref{eq:selection})). $^\ast$Significant difference with the reference (first line), according to a Delong test on the micro-averaged ROC curves ($p<0.05$).}
    \label{tab:ablation_study_2}
\end{table*}

\subsection{Comparison with a 3-D baseline}
\label{sec:baseline}

The proposed DISCOVER framework was also compared to a 3-D baseline model in terms of classification performance and inference times. The baseline model is a 3-D CNN processing the 3-channel 3-D preprocessed acquisition $I$, obtained as presented in section \ref{sec:preprocessing}. We ensured that the same splits were used for training, validation, and testing. Various backbones and hyperparameters were evaluated. The selection process, consistent with the one applied to our proposed method, involved choosing the optimal model based on average AUCs and the best checkpoint for each severity cut-off on the validation set. The most favorable results for the baseline were achieved using a 3-D ResNet50 \citep{he_deep_2016, hara_can_2018} for both 6x6mm$^2$ and 15x15mm$^2$ acquisitions. This baseline and the proposed framework are compared in Table \ref{tab:comparison3Dbaseline}; For the proposed framework, ensemble 1 was used for 6x6mm$^2$ acquisitions and ensemble 2 was used for 15x15mm$^2$ acquisitions, as these are the best ensembles on the validation sets. Classification performances are compared using both Delong tests and a Wilcoxon signed-rank test.

\begin{table*}[!t]
    \centering
    \begin{tabular}{|c|c||c|c|c|c||c|}
        \hline
        \multirow{2}{*}{Acquisition}        & \multirow{2}{*}{Framework} & \multirow{2}{*}{Criterion} & \multirow{2}{*}{AUC} & Delong                         & Wilcoxon                 & Inference times \\
                                            &                            &                            &                      & test ($p$)                     & test ($p$)               & (seconds / volume) \\ 
        \hline
        \hline
        \multirow{8}{*}{6$\times$6mm$^2$}   &                           & $\geq$ mild NPDR            & 0.865                & \multirow{8}{*}{0.1164}        & \multirow{16}{*}{0.0078$^\ast$}
                                                                                                                                                                                                 & \multirow{4}{*}{2.038} \\
                                            & Baseline                  & $\geq$ moderate NPDR        & 0.809                &                                &                                  & \\
                                            & (3-D CNN)                 & $\geq$ severe NPDR          & 0.764                &                                &                                  & \\
                                            &                           & $\geq$ PDR                  & 0.753                &                                &                                  & \\
        \cline{2-4}\cline{7-7}
                                            &                           & $\geq$ mild NPDR            & 0.958                &                                &                                  & \multirow{4}{*}{0.095} \\
                                            & Proposed                  & $\geq$ moderate NPDR        & 0.920                &                                &                                  & \\
                                            & (ensemble 1)              & $\geq$ severe NPDR          & 0.808                &                                &                                  & \\
                                            &                           & $\geq$ PDR                  & 0.821                &                                &                                  & \\
        \cline{1-5}\cline{7-7}
        \multirow{8}{*}{15$\times$15mm$^2$} &                           & $\geq$ mild NPDR            & 0.820                & \multirow{8}{*}{0.0473$^\ast$} &                                  & \multirow{4}{*}{2.876} \\
                                            & Baseline                  & $\geq$ moderate NPDR        & 0.786                &                                &                                  & \\
                                            & (3-D CNN)                 & $\geq$ severe NPDR          & 0.765                &                                &                                  & \\
                                            &                           & $\geq$ PDR                  & 0.886                &                                &                                  & \\
        \cline{2-4}\cline{7-7}
                                            &                           & $\geq$ mild NPDR            & 0.941                &                                &                                  & \multirow{4}{*}{0.251} \\
                                            & Proposed                  & $\geq$ moderate NPDR        & 0.822                &                                &                                  & \\
                                            & (ensemble 2)              & $\geq$ severe NPDR          & 0.876                &                                &                                  & \\
                                            &                           & $\geq$ PDR                  & 0.957                &                                &                                  & \\
        \hline
    \end{tabular}
    \caption{Comparison between the proposed DISCOVER framework and a 3-D CNN baseline in terms of classification performance in the test set (AUC) and in terms of inference times. Inference times are given in seconds per volume, excluding preprocessing (see section \ref{sec:preprocessing}), which is common for both frameworks. For information, inference times are 0.130 seconds/volume for ensemble 2 on 6x6mm$^2$ acquisitions and 0.187 seconds/volume for ensemble 1 on 15x15mm$^2$ acquisitions. $^\ast$Significant difference between the baseline and the proposed approach, according to a Delong test on the micro-averaged ROC curves or according to a Wilcoxon signed-rank test on the paired AUC values ($p<0.05$).}
    \label{tab:comparison3Dbaseline}
\end{table*}

\section{Discussion}

We have presented a general 3-D image classification framework, which combines a trainable \tdtd en-face projection step, followed by a 2-D en-face image classification step. It is further complemented by an auxiliary branch that extracts key 2-D cross-sectional slices (B-scans) and classifies them. The main purpose is to summarize 3-D information by complementary 2-D images (en-face and cross-sectional), for improved interpretability. This novel framework was applied to automated DR severity assessment using 3-D Optical Coherence Tomography Angiography (OCTA) acquisitions. This work aligns with our previous works on interpretable or explainable DR severity assessment \citep{quellec_deep_2017,quellec_explain_2021}. But unlike these previous works, which operated on 2-D Color Fundus Photographs (CFP), this work is applied to 3-D OCTA acquisitions. The additional dimension complicates visual feedback to human readers: the relevant information needs to be summarized so that it can be displayed on a 2-D screen or printed in a report. The proposed framework mimics and generalizes how ophthalmologists analyze OCTA acquisitions: en-face projections are often used to inspect the blood flow; cross-sectional views are often used to inspect structural anomalies. 

Besides improved interpretability, by design, we show that this framework guarantees improved classification performance, compared to direct 3-D image classification. This can be explained by the large volume of information in OCTA acquisitions, which complex 3-D neural architectures cannot process efficiently with limited dataset sizes and limited GPU memory capacities. Summarizing also helps for this purpose. In particular, it allows access to a large collection of highly efficient 2-D neural architectures, with ImageNet pre-trained weights.

\subsection{Analysis of 2-D Projections}

Two types of OCTA acquisitions were analyzed in this study: high-resolution acquisitions centered on the macula (6x6mm$^2$) and lower-resolution ultra-widefield acquisitions (15x15mm$^2$). Fig. \ref{fig:6x6_examples1} shows that normal and pathological retinal features are highly visible in en-face projections for 6x6mm$^2$ acquisitions (columns 3 and 4). This figure highlights the benefit of processing each A-scan independently through 1-D convolutions: all details are lost through U-Net-like projections (U-Projection in column 5). The last column also advocates the joint training of multiple 2-D classifiers through the proposed ``model dropout'' mechanism: by training a single classifier (or static ensemble of classifiers), relevant details may be lost, as the unique classifier may focus on a subset of discriminant features and let the \tdtd projector ignore the others, while still obtaining good classification performance (see Table \ref{tab:ablation_study_1}).

Through comparison of Fig.~\ref{fig:6x6_examples1} and \ref{fig:15x15_examples1}, it appears that 15x15mm$^2$ en-face projections offer a reduced level of details. This makes sense, given the reduced resolution of input images (along all three axes). It is also possible that the \tdtd projector was not able to capture details equally well: the problem is not just a reduced resolution of normal and pathological retinal features, but also a reduced contrast between these features and the background. However, Fig.~\ref{fig:15x15_examples1} suggests that large and peripheral pathological features are well captured in 15x15mm$^2$ acquisitions, explaining that advanced DR stages are detected well in those acquisitions (see Table \ref{tab:performance}).

By design, the proposed \tdtd projection operator $\Pi$, which processes each A-scan independently, does not take the context of these A-scans into account. However, the classification branch $C_1$ does: by training $C_1 \circ \Pi$ jointly, $\Pi$ can be trained to extract information allowing localization. For instance, we can see that color in $\Pi$ varies with the retinal thickness (see Fig. \ref{fig:6x6_examples1} and \ref{fig:15x15_examples1}): this could be useful to capture pathological features (like macular edemas), but certainly also to localize them relatively to the normal retinal features. Indeed, the clinicians collaborating in this study affirm that this method markedly enhances the visibility of retinal lesions, as shown in Fig. \ref{fig:evidence}. As it is apparent, pathological features are well preserved and highlighted, while additional details from the B-scans complement the classifier, reinforcing the need for automated B-scan selection. 

To validate the usefulness of proposed \tdtd projections for decision support, the next step will be to compare the classification performance of clinicians when they use these projections for decision support versus when they do not. Although it is clear that the relevant abnormal structures stand out well in these images, the color-coding of retinal features derived from these projections may be disturbing for some clinicians. For instance, Fig. \ref{fig:6x6_examples1} leads to unusual nebula-like images using ensemble 2 (column 4). For such a validation study, it may be useful to reorder color channels before human inspection, for instance, to produce more conventional-looking images. However, these considerations are out of the scope of an artificial intelligence paper.

One important property illustrated in Fig.~\ref{fig:6x6_examples1} and \ref{fig:15x15_examples1} is that the same features (normal anatomical features like blood vessels, or pathological features like DR lesions) have consistent appearances in 2-D projections across patients (i.e., across lines of these figures). However, the appearance of these features clearly depends on the architecture of the projection operator (i.e., their appearance varies across columns of these figures). Similarly, their appearance depends on the weights trained for the projection operator. If one decides to retrain or fine-tune the proposed framework on a larger dataset, it may be beneficial to freeze the projection weights and only retrain the classification ensembles. This will ensure that the proposed 2-D visualization remains standardized.

\subsection{Detailed Analysis of the Framework}

Tab. \ref{tab:performance} demonstrates that the first classification branch ($C_1$), which analyzes the \tdtd projection, contributes the most to the final classification. This suggests that the proposed \tdtd projections contain most of the discriminant information contained in 3-D OCTA acquisitions for the target classification task, which is good news for interpretability purposes: one single 2-D image conveys most of the relevant information. However, we have shown that combining the two branches leads to a significant increase in classification performance ($p=0.00015$, see Table \ref{tab:performance}). The superiority of branch 1 is particularly true in the two-step training schedule, which was adopted: at first, branch 1 is trained independently, to maximize classification performance; branch 2 ($C_2$), which analyzes selected B-scans, is only trained afterward, to be complementary to the frozen branch 1. In other words, branch 2 was not trained to be discriminant on its own.

An ablation study was performed to analyze the benefits of most methodological choices in the proposed design. Table \ref{tab:ablation_study_1} suggests that classification performance is impacted by the complexity of the projector, driven by parameter $\Phi$, but no significant difference was found with more than $\Phi=8$. Next, the U-Net-like projector and the absence of model dropout, which clearly affect the quality of en-face projections, also decrease classification performance significantly. The use of ImageNet pre-trained weights and logit averaging also proved beneficial. Table \ref{tab:ablation_study_2} suggests that the performance of the second classification branch is little dependent on the attribution method used. As for the random selection of B-scans during training, it seems beneficial, but no significant difference was observed. The most influential parameter in Table \ref{tab:ablation_study_2} is the choice between one-step and two-step training (p=0.0004). The two-step approach has the advantage of more stable training; in particular, one instance of training divergence was observed with the one-step approach. This may explain the increased performance with two steps. On the downside, two-step training implies longer training times (by a factor of two, approximately). Therefore, we recommend investigating both approaches when replicating these results.

\subsection{Comparison with Previous Algorithms}

\begin{table*}[!t]
    \centering
    \begin{tabular}{|c||c|c|c|c|c|}
        \hline
        Reference                                   & Dataset                    & Acquisition                          & Method                          & Criterion                   & Performance \\
        \hline
        \hline
        \citet{heisler_ensemble_2020}               & 380 eyes (463 scans)       & 3$\times$3mm$^2$                     & 2-D CNN                         & $\geq$ rDR                  & acc = 0.92 \\
        \hline
        \citet{le_transfer_2020}                    & 177 eyes (177 scans)       & 6$\times$6mm$^2$                     & 2-D CNN                         & $\geq$ mild NPDR            & AUC = 0.97 \\
        \hline
        \citet{andreeva_dr_2020}                    & 113 patients (320 scans)   & 3$\times$3mm$^2$                     & 2-D CNN                         & $\geq$ mild NPDR            & AUC = 0.83 \\
        \hline
        \citet{lo_federated_2021}                   & 700 eyes (700 scans)       & 3$\times$3mm$^2$                     & 2-D CNN                         & $\geq$ rDR                  & AUC = 0.954-0.960 \\
        \hline
        \multirow{4}{*}{\citet{ryu_deep_2021}}      & \multirow{4}{*}{360 scans} & \multirow{2}{*}{3$\times$3mm$^2$}    & \multirow{4}{*}{2-D CNN}        & $\geq$ mild NPDR            & AUC = 0.928-0.960 \\
                                                    &                            &                                      &                                 & $\geq$ rDR                  & AUC = 0.940 \\
        \cline{3-3} \cline{5-6}
                                                    &                            & \multirow{2}{*}{6$\times$6mm$^2$}    &                                 & $\geq$ mild NPDR            & AUC = 0.926-0.967 \\
                                                    &                            &                                      &                                 & $\geq$ rDR                  & AUC = 0.938-0.976 \\
        \hline
        \multirow{4}{*}{\citet{ryu_deep_2022}}      & \multirow{2}{*}{918 eyes}  & \multirow{2}{*}{3$\times$3mm$^2$}    & radiomics                       & \multirow{4}{*}{DR staging} & acc = 0.574 \\
                                                    &                            &                                      & 2-D CNN                         &                             & acc = 0.684 \\
        \cline{2-4} \cline{6-6}
                                                    & \multirow{2}{*}{917 eyes}  & \multirow{2}{*}{6$\times$6mm$^2$}    & radiomics                       &                             & acc = 0.531 \\
                                                    &                            &                                      & 2-D CNN                         &                             & acc = 0.728 \\
        \hline
        \citet{yasser_automated_2022}               & 91 patients                & 3$\times$3mm$^2$                     & 2-D CNN$^\ast$                  & $\geq$ mild NPDR            & acc = 0.944 \\
        \hline
        \multirow{2}{*}{\citet{zang_diabetic_2022}} & \multirow{2}{*}{355 patients (355 scans)} & 
                                                                                   \multirow{2}{*}{3$\times$3mm$^2$}    & \multirow{2}{*}{3-D CNN}        & $\geq$ rDR                  & AUC = 0.96 \\
                                                    &                            &                                      &                                 & $\geq$ vtDR                 & AUC = 0.92 \\
        \hline
        \multirow{2}{*}{\citet{li_diagnosing_2022}} & \multirow{2}{*}{300 scans} & 3$\times$3mm$^2$ or                  & \multirow{2}{*}{2-D CNN$^\ast$} & \multirow{2}{*}{$\geq$ mild NPDR} & 
                                                                                                                                                            \multirow{2}{*}{AUC = 0.92} \\
                                                    &                            & 6$\times$6mm$^2$                     &                                 &                             &  \\
        \hline
        \citet{li_multimodal_2022}                  & 64 patients (151 scans)    & 6$\times$6mm$^2$                     & 3-D CNN                         & $\geq$ PDR                  & AUC = 0.911 \\
        \hline
        \citet{khalili_pour_automated_2023}         & 78 patients / 148 eyes     & 6$\times$6mm$^2$                     & radiomics                       & PDR v.s. NPDR               & acc = 0.85 \\
        \hline
        
        \multirow{4}{*}{\citet{li_3-d_2023}}        & \multirow{4}{*}{432 patients / 801 eyes} &
                                                                                                                        & \multirow{4}{*}{3-D CNN}        & $\geq$ mild NPDR            & AUC = 0.912 \\
                                                    &                            & 6$\times$6mm$^2$ and                 &                                 & $\geq$ moderate NPDR        & AUC = 0.829 \\
                                                    &                            & 15$\times$15mm$^2$                   &                                 & $\geq$ severe NPDR          & AUC = 0.812 \\
                                                    &                            &                                      &                                 & $\geq$ PDR                  & AUC = 0.900 \\
        \hline
        \multirow{8}{*}{Proposed}                   & \multirow{4}{*}{472 patients / 894 eyes} &
                                                                                   \multirow{4}{*}{6$\times$6mm$^2$}    &                                 & $\geq$ mild NPDR            & AUC = 0.958 \\
                                                    &                            &                                      &                                 & $\geq$ moderate NPDR        & AUC = 0.920 \\
                                                    &                            &                                      & \tdtd                           & $\geq$ severe NPDR          & AUC = 0.808 \\
                                                    &                            &                                      & projection                      & $\geq$ PDR                  & AUC = 0.821 \\
        \cline{2-3} \cline{5-6}
                                                    & \multirow{4}{*}{453 patients / 831 eyes} &
                                                                                   \multirow{4}{*}{15$\times$15mm$^2$}  & \&                              & $\geq$ mild NPDR            & AUC = 0.941 \\
                                                    &                            &                                      & 2-D CNN                         & $\geq$ moderate NPDR        & AUC = 0.822 \\
                                                    &                            &                                      &                                 & $\geq$ severe NPDR          & AUC = 0.876 \\
                                                    &                            &                                      &                                 & $\geq$ PDR                  & AUC = 0.957 \\
        \hline
    \end{tabular}
    \caption{Comparison of DR severity assessment methods using OCTA. The following abbreviations are used: rDR = referable DR; vtDR = vision-threatening DR; PDR = proliferative DR; NPDR = non-proliferative DR; acc = accuracy; AUC = area under the ROC curve. $^\ast$Feature segmentations are used as inputs.}
    \label{tab:comparison}
\end{table*}

Although DR severity assessment using OCTA is a recent topic, a few classification results have already been published, using a variety of DR severity cutoffs; those results are reported in Tab. \ref{tab:comparison}. It should be noted that they were generally obtained on small datasets and that data collection often included a data selection process based on image quality. For instance, \citet{ryu_deep_2021} imaged 496 eyes, but only 360 scans were retained for further analysis, indicating a rejection rate of 27\% (or more, if some patients were imaged more than once). The number of images rejected for quality reasons is not always mentioned: for instance, \citet{le_transfer_2020} only indicated a quality threshold. These factors make comparisons between algorithms challenging. However, it appears that the proposed framework allows similar or better classification results than previously published algorithms, regardless of the type of analysis (2-D CNN, 3-D CNN, or radiomics). Two tasks are particularly well addressed by our framework: $\geq$ mild NPDR detection (AUC = 0.958) and $\geq$ PDR detection (AUC = 0.957). It should be noted that, besides results from our team \citep{li_3-d_2023}, on a subset of the presented dataset, this is the first publication about ultra-widefield OCTA acquisitions (15x15mm$^2$) on the topic. This type of acquisition seems promising for detecting advanced DR stages.\\
In order to thoroughly assess the performance of our proposed method, we have also conducted a direct comparison with a baseline 3-D CNN model, on the same test set (see section \ref{sec:baseline}). We have demonstrated the superiority of the proposed DISCOVER framework over that baseline, both in terms of classification performance and in terms of inference times (see Table \ref{tab:comparison3Dbaseline}). In detail, when comparing ROC curves using Delong tests, a significant difference was found for 15x15mm$^2$ acquisitions, but not for 6x6mm$^2$ acquisitions (p=0.1164). However, when looking solely at the AUC, using a Wilcoxon signed-rank test, a significant difference was found overall in favor of DISCOVER. As for inference times, they are up to 20 times faster using DISCOVER. We believe that our proposed pipeline, which employs a \tdtd summarization in conjunction with a 2-D classification, surpasses the performance of direct 3-D classification models, such as the baseline, for several reasons:
\begin{enumerate}
    \item By incorporating the strengths of both en-face projections and cross-sectional slices, our method captures more pertinent information from the OCTA volumes.
    \item Our method features a lighter architecture compared to 3-D neural networks, resulting in a reduced propensity for overfitting and increased adaptability to smaller datasets. The utilization of pretrained 2-D backbones further bolsters this advantage.
    \item The end-to-end training approach allows our pipeline to acquire more discriminative features, thereby enhancing DR severity assessment.
\end{enumerate}

\subsection{Limitations and Future Works}

This study has a few limitations. First, due to long training times and high resource consumption, no cross-validation or advanced hyperparameter optimization strategy (like Bayesian optimization) was adopted, so chances are that hyperparameter optimization is suboptimal. Thus, in the test set, we observe a better performance with $\Phi=32$ initial filters for two classification labels ($\geq$ severe NPDR and $\geq$ PDR), compared to the hyperparameter value $\Phi=8$ selected in the validation set (see Table \ref{tab:ablation_study_1}). A second limitation is that the dataset is limited in size, which impacts both training and performance evaluation. A third limitation is the use, for training and evaluation, of a DR severity scale known to be suboptimal for DR management due to its limited predictive power.

Addressing the latter two points is the purpose of the Evired project\footnote{\url{https://evired.org/}}: we are collecting longitudinal data from thousands of diabetic patients to define a more predictive DR severity scale. The end goal will be to predict the advent of two DR complications in the following year: proliferative DR and DME. Therefore, ultimately, the proposed framework will be trained to solve a novel 2-label prognosis task ($N=2$). The interpretability of the proposed framework will be particularly useful in discovering which features are found to be predictive.

The use of Transformers could also be investigated in future works. CNNs were used in this paper for convenience: they can generally be applied to images of any size without adaptation. Besides, Transformers often require larger datasets, adding to their computational demands. Given our dataset and computational constraints, the use of CNNs thus proved to be a more practical choice, but this may change in the future.

\section{Conclusion}
This work presents a novel framework for 3-D image classification, with improved interpretability and classification performance. This framework is particularly suited to the analysis of OCT/OCTA images in ophthalmology: its usefulness was demonstrated for DR severity assessment. We expect it to be useful for other 3-D medical image classification tasks in the future.\\
Moving forward, the prospective validation of this framework through a clinical study is an essential next step. Such a study would evaluate whether the selected 2-D projections and B-scans that form the core of our approach indeed provide meaningful and actionable information for clinicians. This validation would support our contention that the methodology not only aids in image classification but also contributes significantly to the interpretability of the results, an aspect that is of high importance for doctors involved in diagnostic processes.\\
Moreover, while the proposed approach has demonstrated promising results, the generalizability of our method will need further assessment. Before our framework can be integrated into routine clinical practice, it is vital to evaluate its performance on an independent population. This evaluation would ensure that the framework is robust and reliable across different populations and clinical contexts and is not limited to the specific cases and data sets upon which it has been developed and tested.

\section*{Declaration of Competing Interest}
The authors declare the following financial interests/personal relationships which may be considered as potential competing interests: Pierre Deman, Employee (ADCIS), Laurent Borderie, Employee (Evolucare Technologies), Hugang Ren and Niranchana Mannivanan, Employees (Carl Zeiss Meditec), B\'eatrice Cochener and Ramin Tadayoni, Consultant (Carl Zeiss Meditec).

\section*{Acknowledgments}
This work was funded in part by the French National Research Agency under the ``Investissement d'Avenir'' program (ANR-18-RHUS-0008 - EVIRED project) and under the LabCom program (ANR-19-LCV2-0005 - ADMIRE project).

\balance

\bibliographystyle{model2-names.bst}
\biboptions{authoryear}
\bibliography{3d-2d-projection, attribution, octa, dr, neural-networks}

\end{document}